\documentclass{emulateapj}

\usepackage{hyperref}
\usepackage{amsmath}
\shorttitle{The $r$-process in the various jet-like explosions of MRD-SNe}
\shortauthors{Nishimura, Takiwaki, and Thielemann}

\begin{document}

\title{The $\lowercase {r}$-process nucleosynthesis in the various jet-like explosions
of magnetorotational core-collapse supernovae}
\author{Nobuya Nishimura\altaffilmark{1}, Tomoya Takiwaki\altaffilmark{2,3}, and Friedrich-Karl Thielemann\altaffilmark{4} }

\affil{
	\altaffilmark{1} Astrophysics Group, Keele University, ST5 5BG Keele, UK; n.nishimura@keele.ac.uk\\
	\altaffilmark{2} Astrophysical Big Bang Laboratory, RIKEN, 351-0198 Wako, Japan\\
	\altaffilmark{3} Center for Computational Astrophysics, NAOJ,
	181-8588 Mitaka, Japan\\
	\altaffilmark{4} Department of Physics, University of Basel, CH-4056 Basel, Switzerland
}

\begin{abstract}
	The $r$-process nucleosynthesis in core-collapse supernovae (CC-SNe) is studied,
	with a focus on the explosion scenario induced by rotation and strong magnetic fields.
	Nucleosynthesis calculations are conducted based on magneto-hydrodynamical explosion models
	with a wide range of parameters for initial rotation and magnetic fields.
	The explosion models are classified in two different types:
	i.e., {\it prompt-magnetic-jet} and {\it delayed-magnetic-jet},
	for which the magnetic fields of proto-neutron stars (PNSs) during collapse and the core-bounce
	are strong and comparatively moderate, respectively.
	Following the hydrodynamical trajectories of each explosion model,
	we confirmed that $r$-processes successfully occur in the prompt-magnetic-jets,
	which produce heavy nuclei including actinides.	
	On the other hand, the $r$-process in the delayed-magnetic-jet is suppressed,
	which synthesizes only nuclei up to the second peak ($A \sim 130$).
	Thus, the $r$-process in the delayed-magnetic-jets could explain only
	``weak $r$-process" patterns observed in metal-poor stars rather than the ``main $r$-process",
	represented by the solar abundances.
	Our results imply that core-collapse supernovae
	are possible astronomical sources of heavy $r$-process elements
	if their magnetic fields are strong enough,
        while weaker magnetic explosions may produce ``weak $r$-process'' patterns ($A \lesssim 130$).
        We show the potential importance and necessity of magneto-rotational supernovae
        for explaining the galactic chemical evolution, as well as abundances of $r$-process enhanced metal-poor stars.
        We also examine the effects of the remaining uncertainties in the nature of PNSs
        due to weak interactions that determine the final neutron-richness of ejecta.        
        Additionally, we briefly discuss radioactive isotope yields in primary jets
        (e.g., $^{56}$Ni),
        with relation to several optical observation of SNe
        and relevant high-energy astronomical phenomena.
\end{abstract}


\maketitle

% -----------------------------------------------------------------------------------------------
\section{Introduction}

The $r$-process---the nucleosynthesis process of rapid neutron capture compared to
$\beta$-decay---is considered to be a main astronomical source of nuclei heavier than iron,
including rare-Earth elements and actinides.
Astrophysical sites for the $r$-process,
which provide suitable neutron-rich ejecta in explosive conditions, are still undetermined,
although a large amount of effort by nuclear physicists, astrophysicists,
and astronomers has been devoted since the dawn of modern science
\citep[for the most recent status, see,][]{2007PhR...450...97A, 2011PrPNP..66..346T}.
The lack of knowledge about the astronomical sources of $r$-process elements
also causes our incomplete understanding
of the galactic chemical evolution of $r$-process elements
and observed abundance patterns of very metal-poor stars.

Core-collapse supernovae (CC-SNe),
which are a main source of heavy elements up to the iron-group nuclei
\citep[e.g.,][]{1995PThPh..94..663H, 1995ApJS..101..181W, 1996ApJ...460..408T,
2002ApJ...576..323R,  2014ApJ...782...91N},
have also been expected to provide reliable conditions for the $r$-process,
because the central remnant object, a neutron star (NS), consists of extreme neutron-rich material.
In this context, the prompt explosion mechanism,
which promptly ejects very neutron-rich matter,
was examined in early studies \citep[see, e.g.,][]{1976A&A....52...63H, 2001ApJ...562..880S}.
However, this explosion scenario itself has been completely excluded
from possible explosion mechanisms by more sophisticated hydrodynamics simulations
with detailed microphysics inputs \citep[][]{2000ApJ...539L..33R, 2001PhRvD..63j3004L}
that employ the Boltzmann neutrino transport and realistic nuclear equations of states (EoSs).

The current standard scenario for CC-SNe
(the so-called ``delayed explosion'') is based on a neutrino-heating mechanism
triggered by convection and hydrodynamical instabilities,
reviving a stalled shockwave \citep{1985ApJ...295...14B}.
Following this scenario, the ejecta appears to be slightly neutron-rich
or even proton-rich in a late phase due to neutrino absorption
\cite[for recent reviews see, e.g.,][]{2012ARNPS..62..407J, 2012AdAst2012E..39K, 2013RvMP...85..245B},
which provides insufficient conditions for producing heavy $r$-process elements.
Recent studies have reported that only some specific cases can lead to successful explosions
avoiding the strong influence of the neutrino absorption.
One candidate is the explosion of lighter massive progenitors ($\lesssim  9 M_{\odot}$),
called the O--Ne--Mg or electron capture SN
\citep[e.g.,][]{2006A&A...450..345K, 2008A&A...485..199J, 2009ApJ...695..208W, 2011ApJ...726L..15W}.
Another possibility is an alternative explosion mechanism
induced by quark-hadron phase transition
\citep[e.g.,][]{2009PhRvL.102h1101S, 2011ApJS..194...39F, 2012ApJ...758....9N}.
These explosion models, however, have shown that only ``weak $r$-processes''
occur in supernova dynamical ejecta,
which mostly produce lighter $r$-process elements ($A \lesssim 130$).

On the other hand, the collapsed core of a massive star, a nascent proto-neutron star (PNS),
has a wind-like outflow from the surface driven by strong neutrino emission.
This outflow is called the PNS wind (or neutrino-driven wind)
and has been regarded as the most reliable astrophysical site for the $r$-process,
since the first proposal and pioneering studies
\citep[e.g.,][]{1992ApJ...399..656M, 1992ApJ...395..202W, 1994A&A...286..857T,
1994A&A...286..841W, 1994A&A...286..857T}.
However, aside from the long-standing problem of high entropies for a successful $r$-process,
the latest hydrodynamics simulations with detailed microphysics have revealed
that the early neutron-rich ejecta of the winds immediately turns proton-rich
\citep{2010A&A...517A..80F, 2010PhRvL.104y1101H,2013JPhG...40a3201A}.
Therefore, it appears insufficient to attain suitable physical conditions for a successful $r$-process
\cite[revisited by][taking into account the recent progress of PNS physics]{2013ApJ...770L..22W}.
However, more advanced treatments of neutrino reactions in dense matter
possibly change the electron fraction ($Y_{\rm e}$) of the winds
\citep{2012PhRvL.109y1104M, 2012PhRvC..86f5803R} and optimistically could lead to a weak $r$-process
\citep[see, also][for the possible influence of neutrino oscillations]{2015PhRvD..91f5016W}.
In conclusion, the PNS wind scenario should be excluded from possible
production sites for heavy $r$-process nuclei,
including the third park elements ($A \sim 195$) and actinides.

Although the PNS wind from canonical CC-SNe hardly produces heavy $r$-process nuclei  ($A \gtrsim 120$),
in case of fast rotation and strong magnetic fields the physical conditions for the $r$-process can change.
Strongly magnetized PNSs, or ``proto-magnetars'',
have been studied as an alternative scenario of regular (in the spherical symmetry) PNS winds
\citep[see, e.g., ][and references therein]{2001PASJ...53..547N, 2004ApJ...611..380T, 2005ApJ...628..914S, 2007ApJ...659..561M}.
Using parameters of rotation and magnetic fields based on several PNS wind models,
more high entropies and/or a rapid expansion of winds seems more favorable
than conditions in non-rotating, unmagnetized PNSs to permit the production of heavy $r$-process nuclei \citep{2014MNRAS.444.3537V}.
Further possible nucleosynthesis signatures of magnetars
(e.g., the production of $^{56}{\rm Ni}$), are discussed in \cite{2015MNRAS.451.4801S}.

Besides CC-SNe and PNS winds, other important candidates are remnant compact binaries,
NS, and NS/black hole (BH) pairs, which eject neutron-rich matter by their collision.
These compact binary mergers (CBMs)
have also been investigated as another possible $r$-process scenario
\citep[e.g.,][]{1974ApJ...192L.145L, 1999ApJ...516..381F, 1999ApJ...525L.121F}.
More sophisticated investigations
\citep[e.g.,][]{2011ApJ...738L..32G, 2012MNRAS.426.1940K, 2013ApJ...773...78B},
have shown that the ejecta of CBMs remains extremely neutron-rich,
producing heavier nuclei than the second $r$-process peak
due to strong neutron capture with fission recycling.
On the other hand, $r$-process elements up to the second peak ($A \sim 130$) in the dynamical ejecta
are severely underproduced,
because fission fragment distributions do not extend to isotopes much below this mass number.
The most recent studies on NS--NS mergers
\citep{2014ApJ...789L..39W, 2015MNRAS.448..541J}
address this open question, based on the shock-heating ejection scenario
with stronger weak interactions (i.e., position captures and neutrino absorption).
Beside this very neutron-rich dynamical ejecta of CBMs,
wind outflows around the central NS and accretion disk
are expected to fill this underproduction of lighter $r$-process elements.
The disk outflow via turbulent and nuclear heating
consists of moderate neutron-rich ($Y_{\rm e} > 0.2$) matter
\citep[e.g.,][]{2009MNRAS.396..304M, 2009MNRAS.396.1659M,
2015MNRAS.448..541J, 2014MNRAS.443.3134P, 2014MNRAS.439..744R}.
Moreover, additional outflows with relatively higher $Y_{\rm e}$ driven by neutrino
are expected for cases of longer duration ($> 100$~ms)
of BH formation from a hyper-massive NS \citep{2014MNRAS.441.3444M}
and the accretion disk of BHs \citep[e.g.,][]{2012ApJ...746..180W}.

The CBM scenario, however, struggles to understand the observational signatures in very early galaxies.
It seems difficult to explain the existence of $r$-process elements
in the early galaxy at the extremely low metallicity of $\rm{[Fe/H]} = - 3$ \citep{2004A&A...416..997A},
because CBMs are delayed due to the formation of binary NSs and the timescale of merging.
This implies that, alternatively, CC-SNe may contribute galactic $r$-process elements
at least in the early universe (at low metallically).
Therefore, potentially multiple astrophysical sites for the $r$-process 
are required, even if CBMs are the major component.
Recently, several studies on galactic chemical evolution
have revisited this problem in different models
with several $r$-process sources, including NS--NS mergers and CC-SNe
\citep[e.g.,][]
{2014A&A...565A..51C, 2014MNRAS.438.2177M, 2014A&A...565L...5T,
2014ApJ...795L..18T, 2015ApJ...804L..35I, 2015MNRAS.447..140V, 2015ApJ...807..115S, 2015MNRAS.452.1970W}.
The full answer to resolve the issue of the role of CBMs and CC-SNe
in the $r$-process enhancement history of the early Galaxy
requires inhomogeneous chemical evolution studies with sufficiently high resolution
to follow the pollution of the interstellar medium by individual events,
which also include all possible mixing processes.

\setcounter{footnote}{4}
CC-SNe induced by strong magnetic fields and/or fast rotation of the stellar core
(i.e., magneto-hydrodynamical supernovae, MR-SNe)\footnote{
Here, this scenario is distinguished from neutrino- and magnetic driven winds from the proto-magnetar,
although they may have some association (i.e., the jet-like explosion of MR-SNe
has been considered to be a birth event of magnetars).},
are considered to provide an alternative and robust astronomical source
for the $r$-process \citep[e.g.,][]{1984ApJ...285..729S, 2003ApJ...587..327C}.
Nucleosynthetic studies were carried out by \cite{2006ApJ...642..410N}
based on adiabatic MHD simulations, which exhibit a successful $r$-process in jet-like explosions.
Additionally, magnetically driven jets of ``collapsar models'' (BH accretion disks)
also have been investigated as a site of the $r$-process
\citep{2006ApJ...644.1040F, 2007ApJ...659..512N, 2009ApJ...704..354H}.
Strong magneto-rotational driven jets of the collapsar model\footnote{
The disk wind of collapsar models were also investigated
\citep[e.g.,][]{2006ApJ...643.1057S, 2008ApJ...679L.117S, 2014JPhG...41d4006S}
as sites of heavy element nucleosynthesis, mostly based on semi-analytic models.}
can produce heavy $r$-process nuclei
\citep{2007ApJ...656..382F, 2008ApJ...680.1350F, 2012PThPh.128..741O}
with a very simple treatment of BH formation.
Here, we should note that the above noted studies for MR-SNe and collapsar models
were performed in the axisymmetric MHD simulations
with simplified treatments of neutrino transport.

One important question is whether the earlier results assuming axis-symmetry
also hold in full three-dimensional (3D) simulations (i.e., lead to the ejection of jets along the polar axis).
MHD simulations in 3D with an improved treatment of neutrino physics
were performed by \cite{2012ApJ...750L..22W} for a $15 M_{\odot}$ progenitor
utilizing the initial dipole magnetic field of $5 \times 10^{12}$~G
and the ratio of magnetic to gravitational binding energy, $E_{\rm mag}/|W| = 2.63 \times 10^{-8}$.
These calculations supported and confirmed the ejection of polar jets in 3D,
attaining $\sim 5 \times 10^{15}$~G and $E_{\rm mag}/|W|=3.02 \times 10^{-4}$ at the core-bounce,
with a successful $r$-process
that produces up to and beyond the third $r$-process peak ($A \sim 195$).
More recent 3D-MHD simulations in the general relativistic framework \citep{2014ApJ...785L..29M},
involving a $25 M_{\odot}$ progenitor with initial magnetic fields of $10^{12}$~G,
initially lead to jet formation,
which afterward experiences a kink instability and deforms the jet-like feature.

Possibly the difference between the two latter investigations in 3D hydrodynamics
marks a transition due to passing critical limits
in the stellar mass and the initial rotation and magnetic fields
between a clear jet-like explosion and a deformed explosion.
Nevertheless, when we see the time evolution of the explosions
with the hydrodynamic instability \citep[see, results in][]{2014ApJ...785L..29M},
the region with the highest magnetic pressure contains essentially the matter
corresponding to the initially forming jets before the deformation,
which is expected to keep the nucleosynthetic features.
For this reason, we justify continuing with the axis-symmetric simulations
instead of 3D-MHD in the current study,
even if the initial jet features will not remain until the completion of the explosion
and the production of $r$-process elements.
We expect that the overall nucleosynthesis composition is close to reality,
which is determined by the neutron-richness of the ejecta
rather than the property of the shock propagation and the timescale of expansion
in the outer layers.

The present study focuses on a remaining problem in the MR-SN scenario
(i.e., the detailed production and ejection mechanism of $r$-process nuclei).
A series of long-term explosion simulations, based on the special relativistic (SR) MHD
\citep{2009ApJ...691.1360T, 2011ApJ...743...30T}, is adopted for the nucleosynthesis examination.
These simulations follow the amplification of magnetic fields due to
differential rotation (winding of magnetic fields) and the launch of the jet-like explosion.
Our MR-SN models include a realistic nuclear EoS, neutrino emission from the PNS,
and a simplified treatment of neutrino transport by the multi-flavor leakage scheme.
Following the studies of \cite{2009ApJ...691.1360T} and \cite{2011ApJ...743...30T},
we classify our explosion models into two categories -- {\it prompt-magnetic-jet}
and {\it delayed-magnetic-jet} explosions --
determined by the strength of the magnetic field just after core-bounce.
We will show that this classification is also applicable for investigating
the production mechanism of $r$-process nuclei.

This paper has the following structure.
Section~\ref{sec-method} briefly summarizes basic physics
and numerical methods for hydrodynamics and nucleosynthesis simulations used in the present study.
The scenario of MR-SNe and the physical properties of explosion models
are explained in Section~\ref{sec-expl}.
The following Section~\ref{sec-res} describes the results of nucleosynthesis, including the $r$-process,
and we also discuss the impacts of the physical uncertainty of MR-SNe on nucleosynthesis in detail.
Finally, Section~\ref{sec-summary} is devoted to summary and conclusions.

% -----------------------------------------------------------------------------------------------
\section{Methods}
\label{sec-method}

We summarize the basic physics and numerical methods of our MR-SN models,
which are based on \cite{2009ApJ...691.1360T}.
Additionally, we describe the TPM,
the treatment of weak interactions at the nuclear statistical equilibrium (NSE)
and relevant physical uncertainties, and the nuclear reaction network.

%................................................................................................
\subsection{Pre-collapse Models}
\label{sec-precol}

We prepare pre-collapse models with rotation and magnetic fields
used in the following hydrodynamical explosion simulations.
Adopting a $25 M_\odot$ pre-collapse model evolved in the spherical symmetry (1D),
we assume simple analytic formula for the distribution of stellar
rotation and magnetic fields at the gravitational collapse,
because the structure and dynamics of the stellar interior in the late phases of evolution
are still poorly known \cite[see, e.g.,][for rotation and magnetic fields]{2002A&A...381..923S}.
The key physical properties of our pre-collapse models are summarized as follows:
 \begin{enumerate}

\item {\it Hydrodynamic structure}---A massive rotating stellar evolution model in the spherical symmetry
\citep[E25 model; described in][]{2000ApJ...528..368H}
is adopted as the standard progenitor in this study.
This is an evolution model that has a $25~M_\odot$ zero-age main-sequence mass
and becomes a Wolf--Rayet star during the core He-burning.
The thermodynamical quantities of the model at the end of evolution
(the pre-collapse stage; i.e., density, internal energy, entropy, and $Y_{\rm e}$)
are used as initial conditions for hydrodynamical simulations.
The enclosed mass and radius of the ``iron core,''
which is composed of iron-group nuclei (e.g., Fe and Ni isotopes),
are $1.69~M_\odot$ and $2188$~km, respectively.

\item {\it Rotation ---}
As commonly used in CC-SN simulations
\citep[see, e.g.,][]{2003ApJ...595..304K, 2004ApJ...616.1086T},
the cylindrical rotation profile is employed for the initial rotation,
which is expressed by the angular velocity in the meridian plane:
\begin{equation}
	\Omega(X,Z) = \Omega_0 \frac{{X_0}^2}{X^2 + {X_0}^2} \frac{{Z_0}^4}{Z^4 + {Z_0}^4}\ ,
	\label{eq-rot}
\end{equation}
where $X$ and $Z$ denote distances from the rotational axis and equatorial plane.
The constant value of $\Omega_0$ represents the strength of rotation,
while $X_0$ and $Z_0$ determine the scale for the region of the fast rotating core
(strength of differential rotation) in the direction of equatorial plane and rotational axis, respectively.
The strength of the rotation is also described by $\beta = E_{\rm{rot}}/|W|$,
which is the ratio of rotational energy ($E_{\rm{rot}}$)
to the absolute value of the gravitational binding energy ($|W|$).
In the present study, the shape of rotation is fixed
by the constants of $X_0 = 100$~km and $Z_0 = 1000$~km,
whereas the strength of rotation varies as
$\Omega_0 = 151$, $76$, and $38$ in rad~s$^{-1}$,
which correspond to $\beta = 0.25 \times 10^{-2}$, $1.00 \times 10^{-2}$, and $4.00 \times 10^{-2}$, respectively.

\item {\it Magnetic fields ---}
Same as rotation, an analytic formula of the initial magnetic fields
is employed for the simplicity.
We assume that the initial magnetic fields have a poloidal dominant structure in the iron core,
which is uniform and parallel to the rotation axis inside the core
and has a dipole shape outside the core.
The shape of the magnetic fields are expressed by an effective vector potential
${\mathbf A}$ in the spherical coordinate $(r,\theta,\phi)$:
\begin{equation}
\begin{split}
	A_r(r,\theta)    &= A_{\theta}(r,\theta) = 0 \ ,\\
	A_\phi(r,\theta) &= \frac{B_0}{2} \frac{{r_0}^3}{r^3 + {r_0}^3}\;r \sin \theta \ ,
	\label{eq-mag}
\end{split}
\end{equation}
which are defined on the meridian $(r,\theta)$-plane.
We choose the radial scale $r_0 = 2000$~km,
which approximately corresponds to the boundary of the iron core.
Additionally, we use two different magnetic field strengths
$B_0 = 1 \times 10^{11}$~G and $1 \times 10^{12}$~G,
which correspond to  $E_{\rm{mag}}/|W| = 2.5 \times 10^{-6}$ and $2.5 \times 10^{-4}$,
respectively.
Here, $E_{\rm{mag}}/|W|$ denotes the ratio of magnetic energy ($E_{\rm mag}$)
to the absolute value of the total gravitational binding energy.
\end{enumerate}

All of the adopted parameters in Equations~(\ref{eq-rot}) and (\ref{eq-mag})
are summarized in Table~\ref{tab-precollapse}.
As shown in this table, we arrange five different initial conditions
for core-collapse and explosion simulations.
In the present paper, each model is named after
its initial quantities of $B_0$ in G and $\beta = E_{\rm rot}/|W|$ (of $10^{-2}$),
(i.e., ``{\tt B11$\beta$0.25},'' ``{\tt B11$\beta$1.00},'' ``{\tt B12$\beta$0.25},''
``{\tt B12$\beta$1.00},'' and ``{\tt B12$\beta$4.00}'').
The explosion scenario of MR-SNe and properties of explosion models
based on the above initial conditions
are described in Section~\ref{sec-expl-scenario} and \ref{sec-expl-model}, respectively.

%%% table 1
\begin{table}[tp]
\centering
\caption{Parameters of Pre-collapse Models}
\begin{center}
\begin{tabular*}{\hsize}{@{\extracolsep{\fill}}lcccc}
\hline
\hline
	Name & $B_0$ & $E_{\rm{mag}}/|W|$ & $\Omega_0$  & $\beta$\\
	& (G)       &      &(rad s$^{-1}$) &  \\
	\hline
	{\tt B11$\beta$0.25} &$10^{11}$ & $2.5 \times 10^{-6}$ & $38$  &$0.25 \times 10^{-2}$  \\
	{\tt B11$\beta$1.00} &$10^{11}$ & $2.5 \times 10^{-6}$ & $76$  & $1.00 \times 10^{-2}$ \\
	{\tt B12$\beta$0.25} &$10^{12}$ & $2.5 \times 10^{-4}$ & $38$  & $0.25 \times 10^{-2}$\\
	{\tt B12$\beta$1.00} &$10^{12}$ & $2.5 \times 10^{-4}$ & $76$  & $1.10 \times 10^{-2}$ \\
	{\tt B12$\beta$4.00} &$10^{12}$ & $2.5 \times 10^{-4}$ & $151$ & $4.00 \times 10^{-2}$ \\
	\hline
\end{tabular*}
\end{center}
\tablecomments
{Adopted parameters in equations (\ref{eq-rot}) and (\ref{eq-mag}),
where $B_0$ is the strength of magnetic fields in G,
$E_{\rm{mag}}/|W|$ is the ratio of magnetic energy to $|W|$,
$\Omega_0$ is the angular velocity in rad~$\rm{s}^{-1}$,
and  $\beta$ is the ratio of rotational energy to $|W|$.
Other scale parameters are assumed as constant:
$X_0 = 100$~km, $Z_0=1000$~km, and $r_0=2000$~km, respectively.}
\label{tab-precollapse}
\end{table}

%................................................................................................
\subsection{Special Relativistic Magnetohydrodynamics}
\label{sec-srmhd}

The nucleosynthesis calculations of the $r$-process, in general,
require long-term hydrodynamical evolution to determine unbound matter
(neutron-rich ejecta).
For this reason, we employ an SR-MHD code,
which has higher computational efficiency and physical validity of dynamics around the speed of light.
Previous investigations based on Newtonian MHD have reported that
the Alfv\'en velocity unphysically exceeds the speed of light
when the jet-head of the shock outflow has strong magnetic fields
\citep[e.g.,][]{2004ApJ...616.1086T}.
The Alfv\'en velocity is simply estimated by
\begin{equation}
	v_{\rm A} = \frac{B}{\sqrt{4\pi\rho}} \sim 10^{10}~\rm{cm}~\rm{s}^{-1}~
	\frac{B/10^{13}~\rm{G}}{\sqrt{\rho/10^5~\rm{g}~\rm{cm}^{-3}}} \ ,
\end{equation}
where $\rho$ and $B$ are the density and the strength of magnetic field, respectively.
If we take typical values around the jet-head of outflow
(i.e., $B \sim 10^{13}$~G and $\rho \sim 10^{5}~{\rm g~cm}^{-3}$),
the second term reaches unity, where $v_A \sim 10^{10}$ $\rm{cm~s}^{-1}$ exceeds the speed of light.
In addition to violating the relativistic causality condition,
this causes the problem of numerical efficiency,
because the maximum $v_{\rm A}$ limits the stable numerical interval of each time step
by the Courant--Friedrichs--Lewy condition.

For this reason, we use an SR-MHD code that has been developed for CC-SN simulations
\citep{2009ApJ...691.1360T, 2011ApJ...743...30T},
based on an Eulerian finite-difference method \citep{1992ApJS...80..753S}.
This SR-MHD code follows the relativistic formalism of \cite{2003ApJ...599.1238D},
where states of relativistic fluid are expressed by hydrodynamic quantities
(i.e., density, $\rho$; internal energy, $e$; velocity, $v^i$; and pressure, $p$)
at each point in the space-time geometry.
This code applies a constrained transport scheme and the method of characteristics
to manage the induction equation and the divergence-free condition, respectively.
The Poisson equation of self gravity is solved by the modified incomplete Cholesky conjugate gradient method.
While solving MHD equations, we also employ a nuclear EoS
\citep{1998NuPhA.637..435S} in the bases of the relativistic mean field theory.
The original EoS table is limited for high densities,
so that we assume that an ideal gas EoS is valid in low densities.
More details on the numerical aspect and computing tests
are available in \cite{2009ApJ...691.1360T} and references therein.

%%% table 2
\begin{table*}[tp]
\centering
\caption{Initial distribution of tracer particles}
\begin{center}
\begin{tabular*}{\hsize}{@{\extracolsep{\fill}}lcccccccccc}
\hline
\hline
	Label
	& $N_{\rm all}$
	& $(N_{r}, N_{\theta})$
	& $\Delta r$
	& $\Delta \theta$
	& $\overline{m}$
	& $m_{\rm min}$
	& $m_{\rm max}$ \\
	&
	&
	& (km)
	& (rad)
	& ($M_\odot$)
	& ($M_\odot$)
	& ($M_\odot$)
	\\
	\hline
	Low & $2500$   & $(50, 50)$ & $79.8$ & $1.54 \times 10^{-2}$ &
	$8.19 \times 10^{-4}$ & $1.22 \times 10^{-4}$ & $2.77 \times 10^{-3}$ \\
	Medium & $10000$  & $(100, 100)$ & $39.9$ & $7.72 \times 10^{-3}$ &
	$2.05 \times 10^{-4}$ & $1.07 \times 10^{-5}$ & $7.00 \times 10^{-4}$ \\
	High & $50000$  & $(500, 100)$ & $7.98$ & $7.72 \times 10^{-3}$ &
	$4.10 \times 10^{-5}$ & $4.34 \times 10^{-7}$ & $1.40 \times 10^{-4}$ \\
	\hline
\end{tabular*}
\end{center}
\tablecomments{$N_{\rm all}$ is the total number of tracer particles.
$N_r$ and $N_\theta$ are the number of particles located in the radial and $\theta$-direction
($N_{\rm all} = N_r \times N_\theta$).
Their intervals are $\Delta r$ and $\Delta \theta$, respectively.
$\overline{m}$ is the average,
$m_{\rm min}$ is the minimum,
and $m_{\rm max}$ is the maximum masses of particles in $M_\odot$.}
\label{tpm-init}
\end{table*}

We perform hydrodynamical simulations on a two-dimensional (2D) computational domain
with the assumption of axisymmetry.
The computational region of the code spreads a quarter of the meridian plane with equatorial symmetry.
We adopt the 2D mesh in the spherical coordinates,
which have $300$ non-uniform grid points in the radial-direction ($r$)
and $60$ uniform points in the polar direction ($\theta$), respectively.
The radial grids cover from $0.5$ to $4000$~km in the radial-direction,
where the innermost grid has the finest interval of $1$~km.
The polar grids constantly spread in the range of $\theta = 0$ -- $\pi/2$ rad,
of which the minimal grid size corresponds to $25$~m.
Additionally, we assume a reflection boundary condition
for both the polar axis and equatorial plane.
The numerical convergence of hydrodynamical simulations was performed
using these 2D special grids on MR-SN simulations,
and is described in Section~5 of \cite{2009ApJ...691.1360T}.

Nucleosynthesis studies on the $r$-process of MR-SNe
\citep{2006ApJ...642..410N, 2012ApJ...750L..22W}
have shown that the $Y_{\rm e}$ of ejecta is influenced
by the electron and positron capture and neutrino absorption in the inner region
($\lesssim 1000$~km) during collapse and explosions.
This implies that weak interactions including neutrino absorption are significant,
as well as for explosion dynamics and formation of the PNS.
However, the neutrino transport has never been solved
in a fully coupled with multi-dimensional hydrodynamics systems,
even by state-of-the-art CC-SN simulations due to computational difficulties.
Thus, we adopt an approximate method for the neutrino cooling of the PNS
by means of the neutrino multi-flavour leakage scheme \citep{2003MNRAS.342..673R}.
This scheme takes into account the change of the $Y_{\rm e}$ by weak-interaction
(i.e., electron and positron captures on nucleons), and photo-, pair-, and plasma-processes
\citep[reaction rates are taken by][]{1978A&A....67..185T, 1985ApJ...293....1F,
1989ApJ...339..354I, 1996ApJS..102..411I}.

The neutrino-spheres of all flavors in which neutrinos are trapped
are consistently defined as the neutrino-optically thick region by the leakage scheme;
whereas, emitted neutrinos from the sphere freely escape outside the surface (the neutrino-optically thin region).
Therefore, when coupled with the leakage scheme, our MHD simulations
ignore the effect of neutrino transport and reactions in the neutrino-optically thin region
(outside the neutrino-sphere),
where the $Y_{\rm e}$ changes due to neutrino absorption.
Including these effects on nucleosynthesis calculations,
we take into account the change of $Y_{\rm e}$
outside the neutrino-sphere within a post-process TPM,
which is described in the following section.

%................................................................................................
\subsection{Tracer Particles with Evolution of $Y_{\rm e}$}
\label{sec-tpm}

The MHD code used in the present study employs grid-based (Eulerian) hydrodynamics,
of which fluid dynamics is given by the time evolution of fluid on spatial numerical mesh.
The particle-based (Lagrangian) hydrodynamics, in contrast,
make use of the initial position (or the particle's number) to describe fluid motion,
which is more reliable to follow the abundance evolution of ejecta.
We adopt the so-called tracer/test particle method (TPM),
which has been used to convert Eulerian multi-dimensional hydrodynamics
into a bulk of Lagrangian particle motion
\citep[see, e.g.,][]{1997ApJ...486.1026N, 2004A&A...425.1029T, 2010MNRAS.407.2297S}.
For this purpose, we employ a TPM code that is an extension of the previous one
\citep[used in][]{2006ApJ...642..410N, 2014ApJ...789L..39W}.

\subsubsection{Numerical schemes}

We briefly summarize the numerical methods (i.e., the time integration and spacial interpolation)
of the TPM code.
The time integration is based on the two variables (2D) predictor-collector method
with the second order accuracy.
At each computational time ($t^n$: the $n$th time step),
we adopt the bilinear interpolation for converting the grid-based physical quantity into tracer particles.
In polar coordinates $(r, \theta)$,
the TPM code calculates the position of the tracer particle
at $t^{n+1}$ (i.e., $(r^{n+1}, \theta^{n+1})$),
using the previous position $(r^n,\theta^n)$
and the velocities in the $r$-direction ($v_r$) and $\theta$-direction ($v_\theta$).
This relation is expressed by
\begin{equation}
	\begin{split}
	r^{n+1}      &= r^n + {{v_r^*}}^n \;\Delta t^n \ ,\\
	\theta^{n+1} &= \theta^n + \frac{{v_\theta^*}^n}{r^n} \;\Delta t^n \ ,
	\end{split}
\end{equation}
where $\Delta t^n (= t^{n+1} - t^n)$, ${v^*_r}^n$, and ${v^*_\theta}^n$
are the time interval, modified-radial-velocity, and modified-angular-velocity, respectively.
Here, ${v_r^*}^n$ and ${{v^*}_\theta}^n$
are functions of velocities at two previous time steps $t^{n-1}$ and $t^n$,
estimated by the predictor-collector scheme.
Additionally, the refraction boundary is assumed for the rotational axis and equatorial plane,
which is the same as the original MHD hydrodynamics simulations.

For the initial position of particles,
we adopt a uniform distribution in space at the beginning of the simulation
(on the pre-collapse model), with different particle masses.
Tracer particles are initially located in a quarter of the meridian plane,
whose domain is $r = 0.5$--$4000$~km and $\theta=0$ -- $\pi/2$ rad.
The innermost initial radius of tracers is $r = 50$~km.
We prepare three different sets of tracer particles in different spatial resolution
for the purpose of a numerical convergence test.
Table~\ref{tpm-init} shows their parameters and relevant physical quantities.
The total number of particles for each set is expressed by $N_{\rm all} = N_{r} \times N_{\theta}$,
where $N_{r}$ and $N_{\theta}$ are the number of particles in the $r$-direction and the $\theta$-direction, respectively.
Additionally, $\overline{m}$, $m_{\rm min}$, and $m_{\rm max}$ are the average,
maximum, and minimum value of masses.
The mass of each tracer particle is conserved during the motion,
of which the initial inner particles have larger values compared to ones in outer layers.
The particle sets are labeled {\it low}, {\it medium}, and {\it high}
expressing their numerical resolution.
The numerical convergence of our TPM estimation is discussed
in Sections~\ref{sec-expl} and \ref{sec-res},
focusing on the $Y_{\rm e}$ of tracer particles and $r$-process nucleosynthesis,
respectively.

%%%%% Fig. 1 Motion of tracer particle (tracer particle method)
\begin{figure}[tbp]
	\begin{center}
		\includegraphics[width=0.9\hsize]{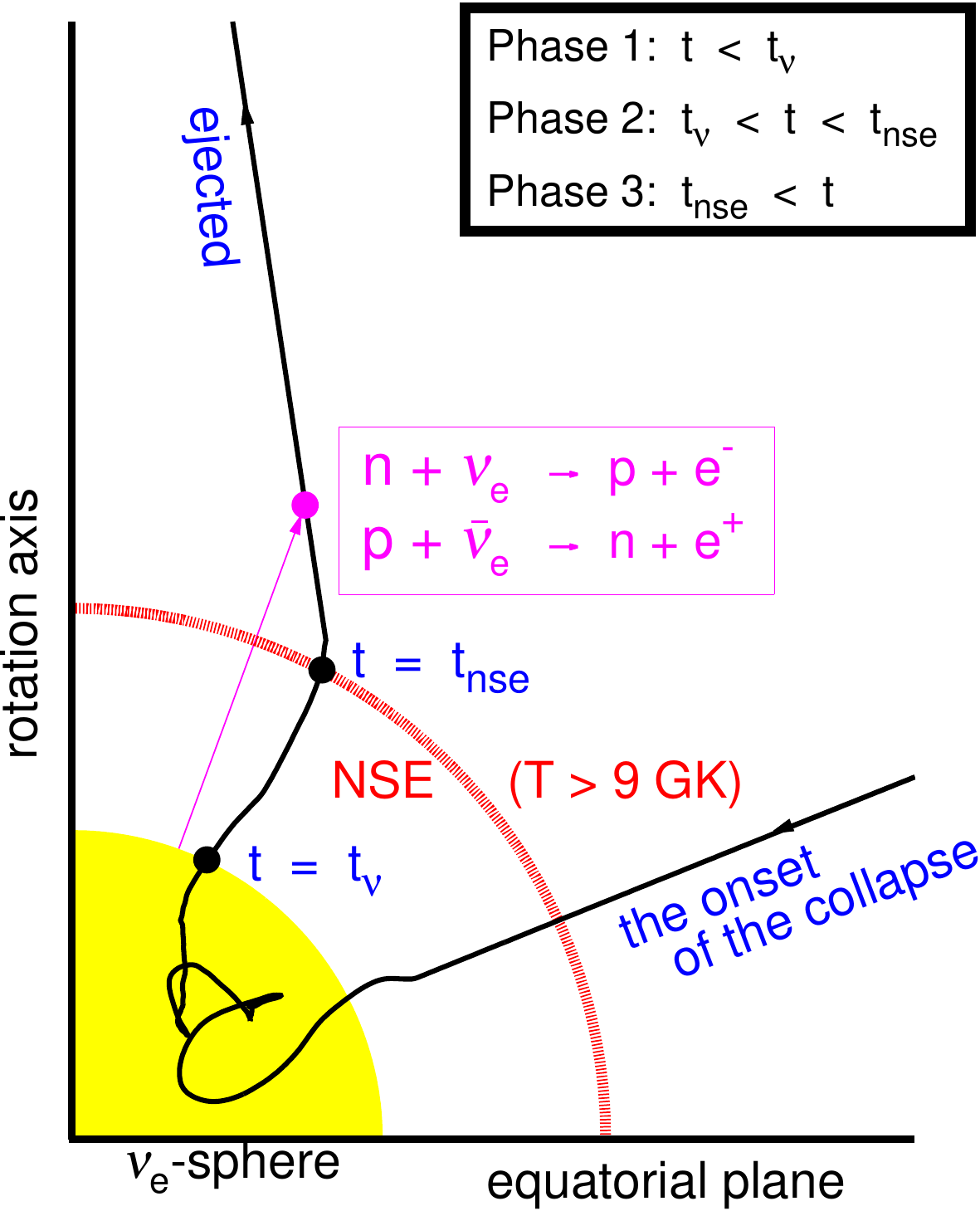}
		\caption{
		A schematic picture of tracer particle motion.
		On the meridian plane, the tracer particle
		that is initially located outside of the core (lower right)
		is finally ejected along the rotation axis (upper left)
		via the convective motion inside the neutrino-sphere (lower left).
		During the motion, the particle reaches the edge of the neutrino-sphere ($t=t_\nu$)
		and escapes form the high temperature NSE region ($t=t_{\rm nse}$).
		}
	\label{fig-n-sphere}
	\end{center}
\end{figure}

\subsubsection{The Evolution of $Y_{\rm e}$ Based on the TPM}
\label{sec-tpm-ye}

The physical value of a tracer particle basically follows the evolution of original hydrodynamical model
that neglects the effect of neutrino absorption outside the neutrino-sphere
(described in Section~\ref{sec-srmhd}).
These neutrino reactions, however, may be crucial for determination of the final $Y_{\rm e}$,
which is important for the resulting $r$-process abundances.
Thus, we will include neutrino-absorption reactions on the TPM during expansion.
For this purpose, we define times from the onset of core-collapse
(i.e., $t_{\nu}$ and $t_{\rm nse}$) for each particle,
focusing on the moment of escape from the neutrino-sphere and NSE state.
Here, $t_{\nu}$ is the moment just before the particle escapes from the neutrino-sphere
and $t_{\rm nse}$ is the time at the end of NSE
that the temperature drops to $9 \times 10^9~{\rm K}$ ($= 9~{\rm GK}$).

Figure~\ref{fig-n-sphere} provides a basic picture of the ejection process
for a typical tracer particle with corresponding time variables ($t_{\nu}$ and $t_{\rm nse}$).
We can safely assume $t_{\rm nse} > t_{\nu}$ for all particles because of the property
of our explosion models, that is, the temperature in the neutrino-sphere is high enough for the NSE,
which agrees with general features of CC-SNe.
Using these two specific times as boundaries,
the ejection history of a tracer is divided into the following three phases:
\begin{enumerate}
\item {\it Phase~1} ($t < t_{\rm \nu}$):
This is the stage that the central core contracts
and the surrounding matter falls, triggered by the gravitational collapse.
The temperature exceeds $1$~MeV ($\sim 11.6$~GK),
where heavy isotopes are destroyed into free nucleons by photo-dissociations
(the NSE state achieves).
On the other hand, densities in the inner core also increase
and become high enough to trap neutrinos inside.
In this phase, we calculate the $Y_{\rm e}$ evolution of tracer particles,
adopting the original hydrodynamical explosion models in the weak equilibrium
based on the Leakage scheme.

\item {\it Phase~2} 
($t_{\nu}<t<t_{\rm nse}$):
The tracer particle has already escaped from the neutrino-sphere,
but the temperature is still high enough ($T > 9$~GK) for the NSE.
Although hydrodynamic simulations ignore neutrino absorption,
we include the effect on changing $Y_{\rm e}$ by means of a post-process treatment.
Using the geometric property of the tracer particle and neutrino-sphere,
we take into account electron and anti-electron neutrino captures on nucleons.
The reaction rates of neutrino absorption are determined by the luminosity,
the mean energy, and the radius of neutrino-spheres.
The isotopic abundances are obtained by solving the NSE (nuclear Saha) equation
that is a function of the temperature and electron density
(determined by the matter density and $Y_{\rm e}$).

\item {\it Phase~3} ($t_{\rm nse} < t$):
After the temperature of the tracer particles drops to $9$~GK,
the NSE becomes invalid,
where all relevant nuclear reactions are activated.
At this stage, we calculate the abundance evolution using a full nuclear reaction network.
The network calculation follows all relevant nuclear reactions
involved in the the $r$-process and other explosive nucleosynthesis,
along with thermodynamical evolution of the tracer particle.
\end{enumerate}

As described, we simply adopt the $Y_{\rm e}$
from hydrodynamics simulations in the neutrino-optically thick region
in the weak equilibrium from the onset of the core-collapse to $t = t_\nu$ (Phase~1).
For the optically thin region (Phases~2 and 3),
where tracer particles escaped from the neutrino-sphere,
we consider the following weak interactions for nucleons:
\begin{equation}
\begin{split}
      {\rm p} + {\rm e^-}  &\rightleftharpoons {\rm n} + {\rm \nu_{\rm e}} \ ,\\
      {\rm n} + {\rm e^+} &\rightleftharpoons {\rm p} + {\rm \bar{\nu_{\rm e}}} \ ,
      \label{weak-reac}
\end{split}
\end{equation}
where p, n, ${\rm \nu_{\rm e}}$, and ${\rm \bar{\nu_e}}$
represent protons, neutrons, electron neutrinos, and anti-electron neutrinos,
respectively.
While the matter is still in the NSE (the temperature exceeds $9$~GK),
we calculate the time evolution of $Y_{\rm e}$, described by the equation:
\begin{equation}
	\frac{d Y_{\rm e}}{dt}
	= - ({\lambda_{\rm pe^-} + \lambda_{\rm p\bar{\nu_{\rm e}}}}) Y_{\rm e}
	+ ({\lambda_{\rm ne^+} + \lambda_{\rm n\nu_{\rm e}}}) (1 - Y_{\rm e}) \ ,
     \label{equ-ye}
\end{equation}
where
$\lambda_{\rm pe^-}$, $\lambda_{\rm ne^+}$,
$\lambda_{\rm n\nu_{\rm e}}$, and $\lambda_{\rm p\bar{\nu_{\rm e}}}$
denote forward and reverse reaction rates in Equation~({\ref{weak-reac}}),
determined by a couple of density and temperature.
We adopt reaction rates for electron and positron captures
from \cite{2001ADNDT..79....1L} given in a grid of temperatures and electron densities.
For their inverse reactions, i.e., neutrino absorptions,
simple analytic formulae \citep{1996ApJ...471..331Q} are used as
functions of the neutrino-sphere radius and the luminosity and mean energy
of neutrino emission from the PNS.

%%%%% Fig. 2 Heavy elemental abundances at NSE
\begin{figure}[tbp]
	\begin{center}
		\includegraphics[width=\hsize]{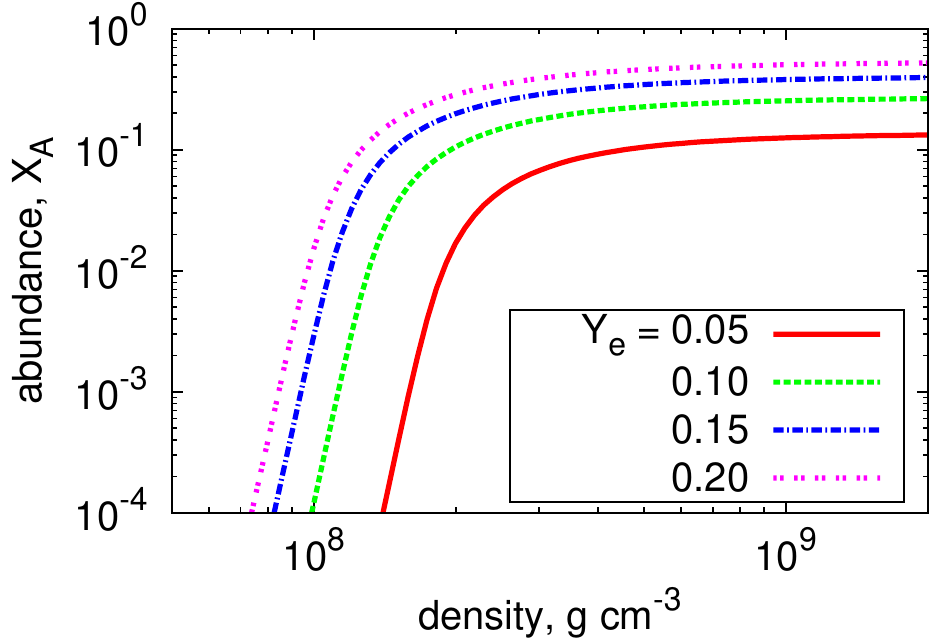}
		\caption{
		Abundances of heavy elements at NSE.
		The total mass fraction of heavy nuclei ($X_A$),
		defined by their mass number, $A \ge 6$,
		are plotted in a typical density range.
		The cases of $Y_{\rm e} = 0.05$, $0.10$, $0.15$, and $0.20$
		are calculated with a constant temperature $T=9$~GK.
		}
	\label{fig-nse}
	\end{center}
\end{figure}

The duration of hydrodynamical simulations ($\sim 100$~ms after the core-bounce)
is insufficient to follow all relevant nucleosynthetic processes, even for the $r$-process.
For the $r$-process, neutron captures and $\beta$-decays continue
several hundred seconds ($\sim1$ -- $100$~s) during expansion.
We obviously do not have the time evolution of hydrodynamical values based on MHD simulations
in the phase of $r$-process nucleosynthesis.
However, in such a late stage of the explosion,
the dependence of the hydrodynamic evolution (the expansion timescale)
on nucleosynthesis becomes comparably weak,
so that we can safely assume a very simple expansion law.
Additionally, the most important variable $Y_{\rm e}$,
which determines neutron-richness,
immediately reaches the asymptotic value after the NSE\footnote{
We will discuss the evolution of $Y_{\rm e}$ and other physical values in Section~\ref{sec-traj-evol}.
Figure~\ref{fig-ye-evol} shows a typical time evolution of $Y_{\rm e}$
for selected trajectories.} (Phase~3),
although the $r$-process itself changes $Y_{\rm e}$ due to $\beta$-decay
in later stages ($t > 100$~ms after the bounce).

Here, we use an analytic extrapolation to fill the gap
between the time duration of the simulation and the need for nucleosynthesis simulations.
Assuming adiabatic expansion after the end of the simulations ($t_{\rm f}$),
we have the constant velocity $v(t) = v_{\rm f}$,
radius $r(t) = r_{\rm f} + (t - t_{\rm f})\,v_{\rm f}$,
density $\rho(t)$, and temperature $T(t)$, described by
\begin{equation}
	\begin{split}
	\rho(t) &= \rho_{\rm f} \left[ \frac{r_{\rm f}}{r(t)} \right]^3,\\
	T(t)    &= T_{\rm f} \frac{r_{\rm f}}{r(t)} \ ,
	\end{split}
\end{equation}
where $t_{\rm f}$, $v_{\rm f} = v(t_{\rm f})$, $r_{\rm f}$, $\rho_{\rm f}$,
and $T_{\rm f}$ are the final velocity, radius, density, and temperature
of the hydrodynamical simulation, respectively.
Although this extrapolation may not be realistic for modeling the hydrodynamics of explosion models,
the impacts on nucleosynthesis by physical uncertainties are expected to be negligible
because our hydrodynamical simulations follow time evolution at least until the end of NSE,
satisfying $t_{\rm f} > t_{\rm nse}$ \citep[see,][]{2006ApJ...642..410N}.
On the other hand, hydrodynamical instabilities may change the dynamics
in the early stage of Phase~3 ($\sim 100$~ms),
which launched jets that are deformed due to the kink instability \citep{2014ApJ...785L..29M}
and experienced thermal energy deposition by magnetic reconnection.
We quantitatively discuss the impacts of the kink instability on the $r$-process
in Section~\ref{sec-jet-kink}.

% table 3
\begin{table}[tp]
\centering
\caption{Properties of the neutrino-sphere and emission}
\begin{center}
\begin{tabular*}{\hsize}{@{\extracolsep{\fill}}c|cccccc}
\hline
\hline
	Model
	& $R_{\rm \nu_{\rm e}}$   & $R_{\rm \bar{\nu_{\rm e}}}$
	& $L_{\rm \nu_{\rm e}}$ & $L_{\rm \bar{\nu_{\rm e}}}$\\
	& (km) & (km) & ($10^{52}$ erg s${}^{-1}$) & ($10^{51}$ erg s${}^{-1}$)\\
	\hline
	{\tt B11$\beta$0.25} & $94.0$ & $79.5$ & $4.33$ & $4.76$\\
	\hline
	{\tt B11$\beta$1.00} & $102$  & $79.6$ & $4.76$ & $1.77$\\
	{\tt B12$\beta$0.25} & $98.6$ & $76.6$ & $3.87$ & $3.48$\\
	{\tt B12$\beta$1.00} & $101$  & $78.4$ & $3.69$ & $1.53$ \\
	{\tt B12$\beta$4.00} & $104$  & $80.0$ & $3.52$ & $6.78$ \\
	\hline
\end{tabular*}
\end{center}
\tablecomments
	{The radiuses of neutrino-sphere; $R_{\nu_{\rm e}}$ and $R_{\bar{\nu_{\rm e}}}$
	for the electron neutrino and anti-electron neutrino, respectively.
	The luminosities of neutrino emission; $L_{\nu_{\rm e}}$ and $L_{\bar{\nu_{\rm e}}}$
	after the neutrino burst for the electron neutrino and anti-electron neutrino.}
\label{tab-neutrino}
\end{table}

%................................................................................................
\subsection{Nuclear Statistical Equilibrium}
\label{sec-tpm-nse}

During early stages of the explosion (Phase~1~and~2; in Figure~\ref{fig-n-sphere}),
most of ejecta are in the state of NSE ($T>9$~GK), of which isotopic abundances
are immediately determined by the temperature and electron number density
(the product of the density and $Y_{\rm e}$).
If the ejecta escaped from the weak equilibrium region (inside the neutrino-sphere),
the value of $Y_{\rm e}$ changes due to weak interactions despite being in the NSE,
in which all nuclear reactions balance with their inverse reactions.
In the present study, we assume that the NSE is valid
if the temperature exceeds $9$ GK for any electron number densities.
Above this temperature, we calculate NSE isotopic abundances
by solving the NSE (nuclear Saha) equation with weak interactions
\citep[see,][]{2012ApJ...758....9N}.

By solving the NSE equation,
we obtain isotopic abundances
at a given set of the temperature, density, and $Y_{\rm e}$.
Figure~\ref{fig-nse} shows the total mass fraction of heavy elements ($X_A$) in NSE,
which is the summation of heavy ($A \ge 6$) isotopes.
We calculated abundances in different values of $Y_{\rm e}$,
with a constant temperature, $9$~GK, the lower bound of the NSE.
The chosen range of densities and $Y_{\rm e}$ covers
typical values of neutron-rich ejecta from MR-SNe,
where the density $\rho < 10^{10}$ g $\rm{cm}^{-3}$
and $Y_{\rm e} = 0.05, 0.10, 0.15$, and $0.20$.
As shown in the figure, heavy elements are negligible
when densities are below $\sim 10^8$ g $\rm{cm}^{-3}$
for any $Y_{\rm e}$.
At high densities with low $Y_{\rm e}$,
the values of $X_A$ are still less than $\sim 10$\,\%,
where $\rho \geq 10^9 $ g ${\rm cm}^{-3}$ with $Y_{\rm e} \sim 0.05$ -- $0.10$, respectively.
Therefore, we can assume that heavy elements have minor effects on weak interactions.

In addition to electron and positron captures on nucleons,
neutrino absorption is taken into account outside the neutrino-sphere
(Phase~2 and 3; in Figure~\ref{fig-n-sphere}).
We use the absorption rates of \cite{1996ApJ...471..331Q},
determined by the distance from the surface of a neutrino-sphere
and the physical properties of neutrino emission from the PNS,
(i.e., the mean energy and luminosity).
These relevant quantities are consistent with the hydrodynamical explosion models,
which are simulated with a simplified treatment of the neutrino transport by the Leakage scheme.
The adopted values of the radius of neutrino-sphere and the luminosity of neutrino emission
are summarized in Table~\ref{tab-neutrino} for all explosion models,
whose uncertainties are discussed in more detail in Section~\ref{sec-ye}.
Using these reaction rates,
Equation~(\ref{equ-ye}) is solved to calculate the evolution of $Y_{\rm e}$ for each tracer particle
along the time evolution of the temperature and density.

%................................................................................................
\subsection{Nuclear Reaction Networks}
\label{sec-network}

We utilize a full nuclear reaction network,
which consists of all relevant isotopes and reactions to the $r$-process,
for nucleosynthesis calculations.
We use an extension of the previous code,
which has been described in detail
\citep[see][]{2012PhRvC..85d8801N, 2012ApJ...758....9N}.
This network consists of $4423$ isotopes from the proton and neutron
up to fermium (which has an atomic number $Z = 100$),
which includes proton-rich isotopes as well as neutron-rich ones far from $\beta$-stability isotopes.
This reaction network includes decay channels, two- and three-body reactions,
electron and positron captures on nuclei, and screening effects,
as shown in the following equation for the specific $i$th isotope, denoted by $Y_i$,
that is
\begin{equation}
\begin{split}
	\frac{dY_i}{dt}
	=& \sum_j {{\mathcal N}_{ij}} \,\lambda_j \,Y_j \\
	&+ \sum_{j,\,k} {\mathcal N}_{i,\,j,\,k} \,\rho N_{\rm Av} \,\langle j, k \rangle \, Y_j \,Y_k \\
	&+  \sum_{j,\,k,\,l} {\mathcal N}_{i,\,j,\,k,\,l} \,\rho^2 {N_{\rm Av}}^2
	\,\langle j, k, l \rangle \, Y_j \,Y_k \,Y_l \ ,
\end{split}
\label{eq-network}
\end{equation}
where $N_{\rm Av}$ is Avogadro's number, and
${\mathcal N}_{i,\,j}$, ${\mathcal N}_{i,\,j,\,k}$, and ${\mathcal N}_{i,\,j,\,k,\,l}$,
are proper accounting numbers avoiding overlapped-counting, defined by
${\mathcal N}_{i,\,j}         = N_i$,
${\mathcal N}_{i,\,j,\,k}     = N_i /\Pi^{n_m}_{m=1} |N_{j_m}|!$, and
${\mathcal N}_{i,\,j,\,k,\,l} = N_i /\Pi^{n_m}_{m=1} |N_{j_m}|!$,
respectively.
$N_{i}$ is the factor that denotes the variation of
the $i$th isotope in all decays and reactions.
Additionally, $\lambda_i$ is the decay constant,
and $\langle j, k \rangle$ and $\langle j, k, l \rangle$ are the cross-sections
\citep[for details, see,][]{1999JCoAM.109..321H}.

At each time step, our network code solves a large linear equation system,
which is based on Equation~(\ref{eq-network}),
by matrix inversion using a sparse matrix solver, UMFPACK\footnote{
http://faculty.cse.tamu.edu/davis/suitesparse.html}
\citep{Davis:2004:AUV:992200.992206}.
We use a semi-implicit method coupled with the Newton--Raphson method for the time integration.
Each thermonuclear reaction rate is given as a function of the temperature and density,
where the thermodynamical evolution of a tracer particle is adopted.
We choose experimental-based nuclear masses
provided by a series of experimental nuclear mass evaluation
\citep{1995NuPhA.595..409A}.
Additionally, theoretical predictions for nuclear
masses, reaction rates, and beta-decays are applied, based on
the finite-range droplet model (FRDM) mass formula \citep{1995ADNDT..59..185M}.

The majority of reaction and decay rates have been taken
from the compilation of REACLIB database\footnote{http://nucastro.org}
\citep{2000ADNDT..75....1R} and its extension, JINA REACLIB\footnote{
https://groups.nscl.msu.edu/jina/reaclib/db}
\citep{2010ApJS..189..240C},
in which theoretical ones are mainly based on FRDM.
Physical uncertainties of $\beta$-decay half-lives
and the effects of recent experimental results on the $r$-process
in our network code were discussed in \cite{2012PhRvC..85d8801N}.
We partially utilized experimental-based reaction rates
from NACRE \citep{1999NuPhA.656....3A} for lighter nuclei.
For heavy elements that we take into account spontaneous and $\beta$-delayed fission,
we assume an empirical formula of fission fragments \citep{1975NuPhA.239..489K}.
Further details of adopted isotopes, nuclear reactions, and fissions
have been explained in our previous papers \citep{2006ApJ...642..410N, 2008ApJ...680.1350F}.
We take into account weak interactions,
which are significant to determine the proton/nucleon ratio $Y_{\rm e}$,
as well as nuclear reaction rates described above.
For the electron and positron capture by heavy nuclei including iron-group isotopes,
we use tabled reactions \citep{1980ApJS...42..447F, 1982ApJS...48..279F, 2001ADNDT..79....1L}.
Neutrino absorption ($\nu_{\rm e}$ and $\nu_{\bar{\rm e}}$) by nucleons is also taken into account,
which is explained in the previous section.
Unlike other nuclear reactions that are determined
by a combination of the density, temperature, and $Y_{\rm e}$,
neutrino-absorption reactions are also depend on
the distance from the neutrino-sphere and the properties of neutrino emission from the PNS.
We adopt these quantities based on our explosion models,
which is the same manner of estimating $Y_{\rm e}$ during the NSE state outside the neutrino-sphere
(described in Section~\ref{sec-tpm-nse}).

% -----------------------------------------------------------------------------------------------
\section{Magnetorotational Supernovae}
\label{sec-expl}

We describe the explosion scenario of MR-SNe
and the ejection mechanism of neutron-rich matter,
based on our explosion models.
We focus, in particular, on the evolution of $Y_{\rm e}$
that directly relates to the following $r$-process nucleosynthesis.
Additionally, we discuss the uncertainty of $Y_{\rm e}$ in our calculations
due to the physical properties of a PNS.

%%%%% Fig. 3 3D structure of MR-SNe
\begin{figure*}[tbp]
	\begin{center}
		\includegraphics[width=0.425\hsize]{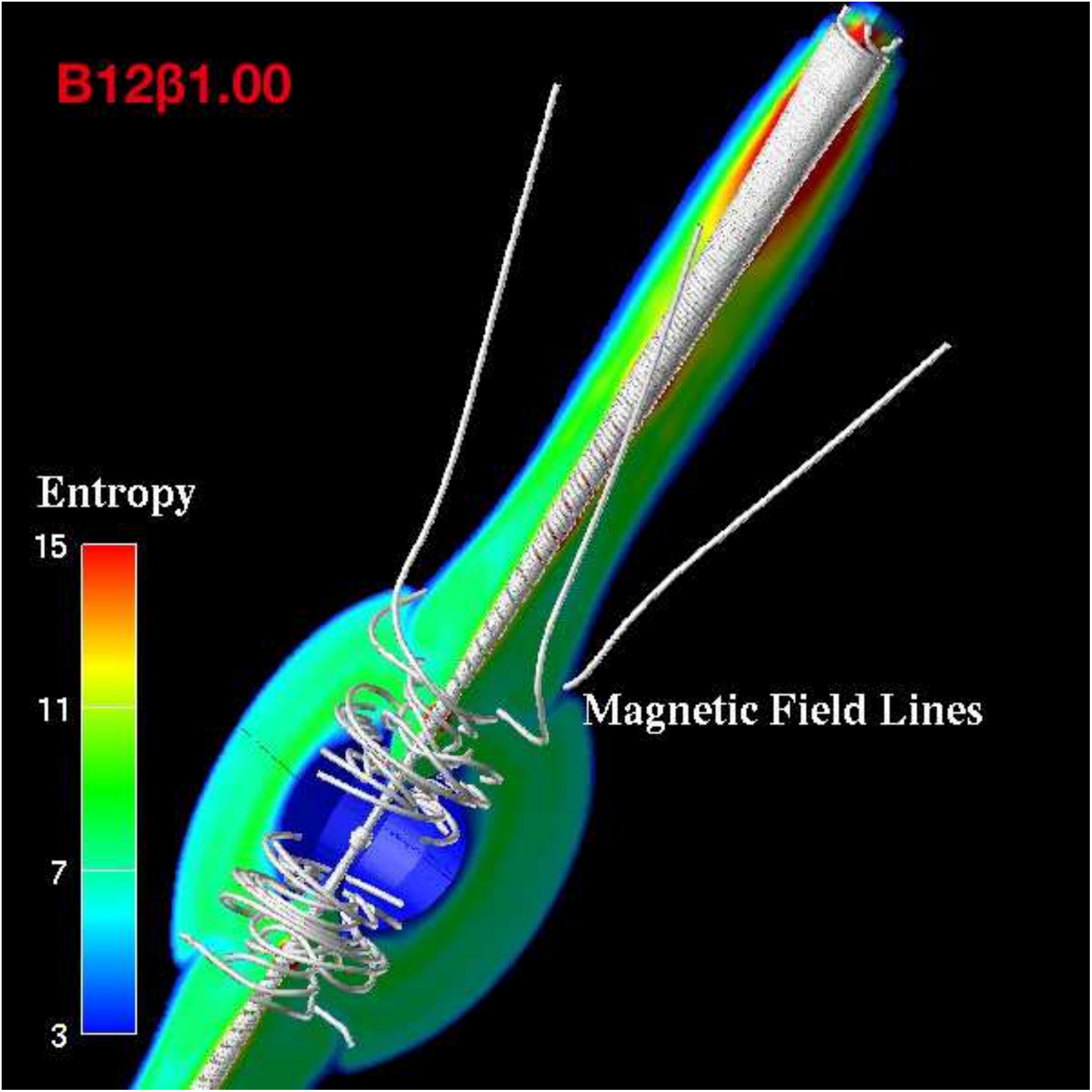}
		\hspace{0.02\hsize}
		\includegraphics[width=0.425\hsize]{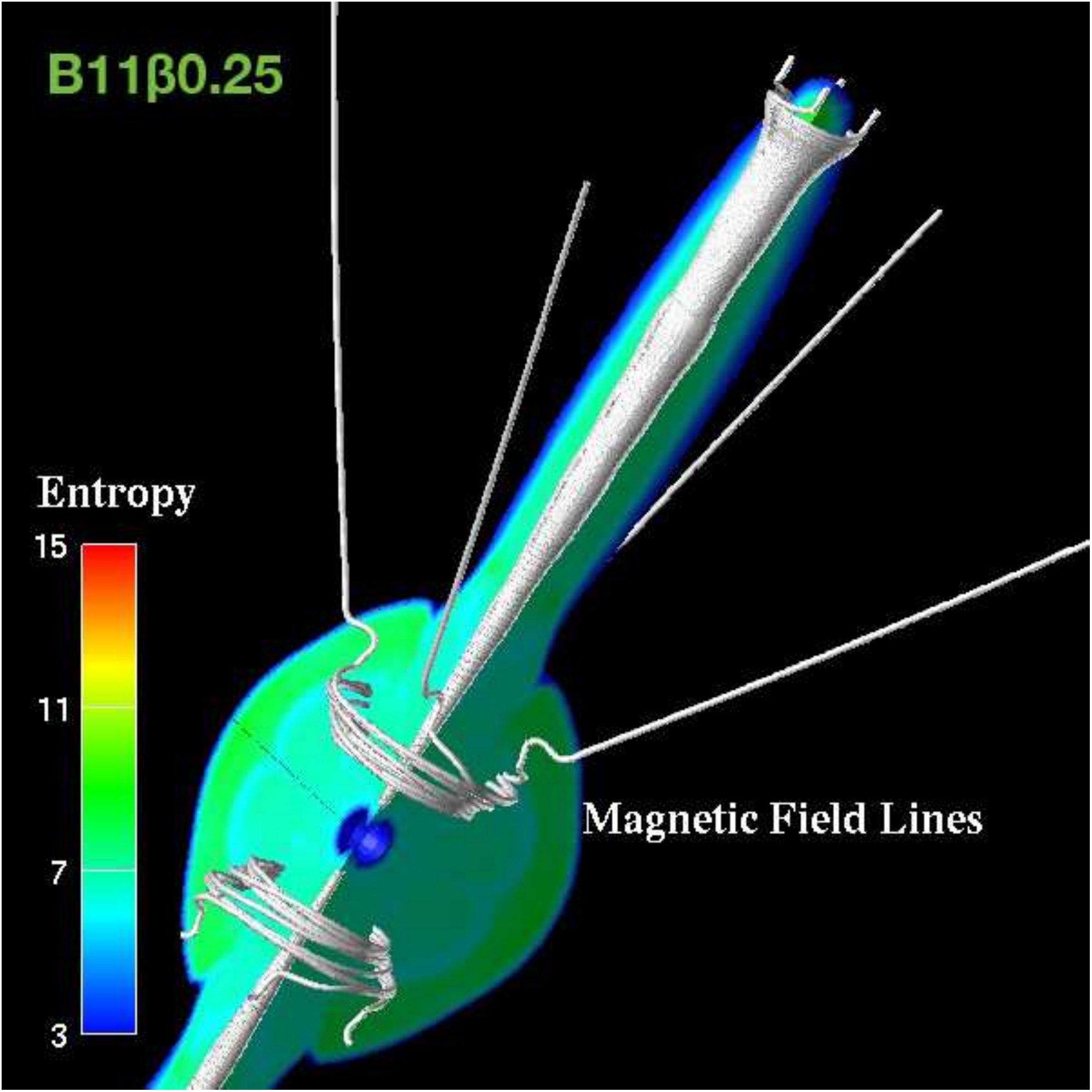}
		\caption{
		The structure of MR-SNe with a bipolar-jet explosion around the PNS.
		The entropy per baryon ($3$--$15$ $k_{\rm B}$ baryon$^{-1}$) and magnetic field lines (white lines)
		are drawn in the spatial range of $700$~km.
		The left and right panels correspond to {\tt B12$\beta$1.00} (prompt-jet model)
		and {\tt B11$\beta$0.25} (delayed-jet model), respectively.}
	\label{fig-3d-jet}
	\end{center}
\end{figure*}
%%% table 4
\begin{table*}[tp]
\centering
\caption{Physical properties of explosion models}
\begin{center}
\begin{tabular*}{\hsize}{@{\extracolsep{\fill}}lcccccccccc}
\hline
\hline
	Name
	& Category
	& $t_{\rm{bnc}}$
	& $t_{\rm{del}}$
	& $t_{\rm exp}$
	& $t_{\rm{fin}}$
	& $v_{\rm{jet}}$
	& $E_{\rm{exp}}$
	& $M_{{\rm ej},r}$
	& $\langle Y_{{\rm e},r} \rangle$
	\\
	&
	& (ms)
	& (ms)
	& (ms)
	& (ms)
	& ($c$)
	& ($10^{50}\,\rm{ergs}$)
	& ($10^{-3}\,M_{\odot}$)
	&
	&
	\\
	\hline
	\hline
	{\tt B11$\beta$0.25}  & delayed & $244$ & $48$ & $72$ & $103$ & $0.12$ & $0.05$ & $2.68$ & $0.281$\\
	\hline
	{\tt B11$\beta$1.00}  & prompt  & $235$ & $10$ & $27$ &  $58$ & $0.27$ & $0.23$ & $1.62$ & $0.192$\\
	{\tt B12$\beta$0.25}  & prompt  & $244$ & $ 0$ & $32$ &  $57$ & $0.20$ & $1.3 $ & $3.69$ & $0.186$\\
	{\tt B12$\beta$1.00}  & prompt  & $235$ & $ 0$ & $20$ &  $77$ & $0.27$ & $1.4 $ & $4.05$ & $0.221$\\
	{\tt B12$\beta$4.00}  & prompt  & $292$ & $ 0$ & $25$ &  $33$ & $0.20$ & $1.0 $ & $5.32$ & $0.181$\\
	\hline
	\hline
\end{tabular*}
\end{center}
\tablecomments{
$t_{\rm bnc}$ is the duration of the collapse, which is from the onset of core-collapse until the core-bounce.
The other three timescale are measured from the core-bounce ($t_{\rm bnc}$):
$t_{\rm del}$ is the raunch of outward shock, 
$t_{\rm exp}$ is the shock reaching the distance of $1000$~km from the PNS,
and $t_{\rm fin}$ is the end of simulation.
The velocity of the shock at $t_{\rm exp}$ ($1000$~km from the PNS)
is $v_{\rm jet}$, with a corresponding explosion energy of $E_{\rm exp}$.
$M_{{\rm ej},\,r}$ and $\langle Y_{{\rm e},\,r} \rangle$
are the mass and the average $Y_{\rm e}$ of very neutron-rich ejecta
($ Y_{\rm e} < 0.3$), respectively.
}
\label{tab-expl-properties}
\end{table*}

\subsection{Explosion scenario}
\label{sec-expl-scenario}

The mechanism of magnetorotational induced SNe (MR-SNe),
associated with a jet-like explosion is clearly distinguished
from the neutrino-heating mechanism of canonical CC-SNe.
In principle, the dynamics of fast rotating iron cores has common properties
from the onset of gravitational collapse until the core-bounce,\footnote{
The gravitational collapse of the iron core (formation of the PNS)
is triggered by the same physical processes as standard CC-SNe,
which are the photo-dissociation and electron capture.}
even if the core is significantly magnetized
\citep[see,][and references therein]{2003ApJ...595..304K}.
The strength of the initial rotation in an iron core, which is determined by $\beta = T/|W|$,
governs dynamics before the core-bounce,
where the angular momentum transfer is ignorable \citep{2004ApJ...608..907Y}.
Strong magnetic fields in the PNS become significant for explosion dynamics after the core-bounce.

The dynamics of MR-SNe, which is driven by rapid rotation and strong magnetic fields,
may have two different types, characterized by the amplification process
of the magnetic fields around a PNS.
Following the study of \cite{2009ApJ...691.1360T},
we classify our explosion models into two categories:
the {\it prompt-magnetic-jet} and {\it delayed-magnetic-jet} explosions,
or simply the {\it prompt-jet} and {\it delayed-jet}, respectively.
The former type, prompt-jets, promptly explodes just after the core-bounce
associated with an energetic bipolar-jet explosion,
due to a strong magnetic pressure by rapid rotation and strong magnetic fields at the pre-bounce stage;
the latter case, delayed-jets, is a relatively weak explosion
with a more collimated jet-like explosion.
The delayed-jets have moderate initial rotation and magnetic fields compared to the prompt-jets,
and long-term duration until the launch of a jet-like outflow after the core-bounce.
In the following we summarize the characteristic properties of these two types for MR-SN models;
pictures of 3D hydrodynamical structure (in axisymmetry) are shown in Figure~\ref{fig-3d-jet}.

{\it Prompt-magnetic-jets.}---The pre-collapse core has strong initial magnetic fields with rapid rotation,
and the wound-up process during the collapse enhances the magnetic fields around the central region.\footnote{
The ``plasma-$\beta$'' (this is not $\beta=T/|W|$),
which is the ratio of the matter pressure to the magnetic pressure,
decreases to $10^{-2}$ in this region \citep{2009ApJ...691.1360T}.}
Strong toroidal magnetic fields above $10^{15}$~G
around the surface of PNS launch a strong shockwave
that overcomes the ram pressure of falling matter.
This jet-like shockwave runs along the rotational axis, so that the outer layers finally expload in the polar direction;
whereas, shock propagation in the equatorial plane immediately diminishes and fails to explode.
Only an energetic bipolar-jet explosion, which is collimated by strong magnetic fields
(the left panel of Figure~\ref{fig-3d-jet}), propagates.
As shown in the figure, the maximum value of entropies inside the jet
reaches $\sim 15~k_{\rm B}~{\rm baryon}^{-1}$;
the central region has lower values ($\sim 5~k_{\rm B}~{\rm baryon}^{-1}$).

{\it Delayed-magnetic-jets.}---This explosion mechanism is caused by comparatively weak initial rotation and magnetic fields,
which takes longer duration from the core-bounce for a successful explosion.
After the core-bounce, the outward shockwave in the direction of the rotational axis stalls,
as does the the shockwave in the equatorial plane,
because of a weaker magnetic pressure around a PNS.
This stalled shock in all directions begins to oscillate until $\sim 10$--$100$~ms after the bounce.
Although this oscillation is diminishing and reaches a nearly hydrostatic state,
the magnetic fields behind the shock grow due to magnetic fields wrapping,
where the plasma-$\beta$ decreases
(i.e, the magnetic pressure becomes predominant).
When the toroidal component of magnetic fields exceeds $10^{15}$~G,
the stalled shockwave revives and propagates along the rotational axis.
Comparing to the prompt-jets,
the shape of the delayed-jets is more collimated but less energetic.
As shown in the right panel of Figure~\ref{fig-3d-jet},
magnetic field lines are strongly wrapped around the PNS.

In the present study, we investigate nucleosynthetic properties of the MR-SN,
mainly focusing on the mechanism of explosion and the ejection process of neutron-rich material.
Investigations of the MR-SN are also an important subject of numerous unsolved problems
(i.e., macroscopic and microscopic amplification processes of magnetic fields,
neutrino bursts, gravitational wave detection, energetic supernova events,
and the study of GRBs).
The predicted properties of gravitational wave signals and neutrino bursts
are apparently different between the prompt- and delayed-jets \citep{2011ApJ...743...30T}.
Additionally, the effects of anisotropic neutrino bursts and influences to the neutrino-oscillation
via the jet-like explosion have been investigated by \cite{2009JCAP...09..033K}.
For more details of those several studies and relevant issues to MR-SNe,
see \cite{2012AdAst2012E..39K} and references therein.

%%%%% Fig. 4 Jet evolution by tracer particles
\begin{figure*}[tbp]
	\begin{center}
		\includegraphics[width=0.95\hsize]{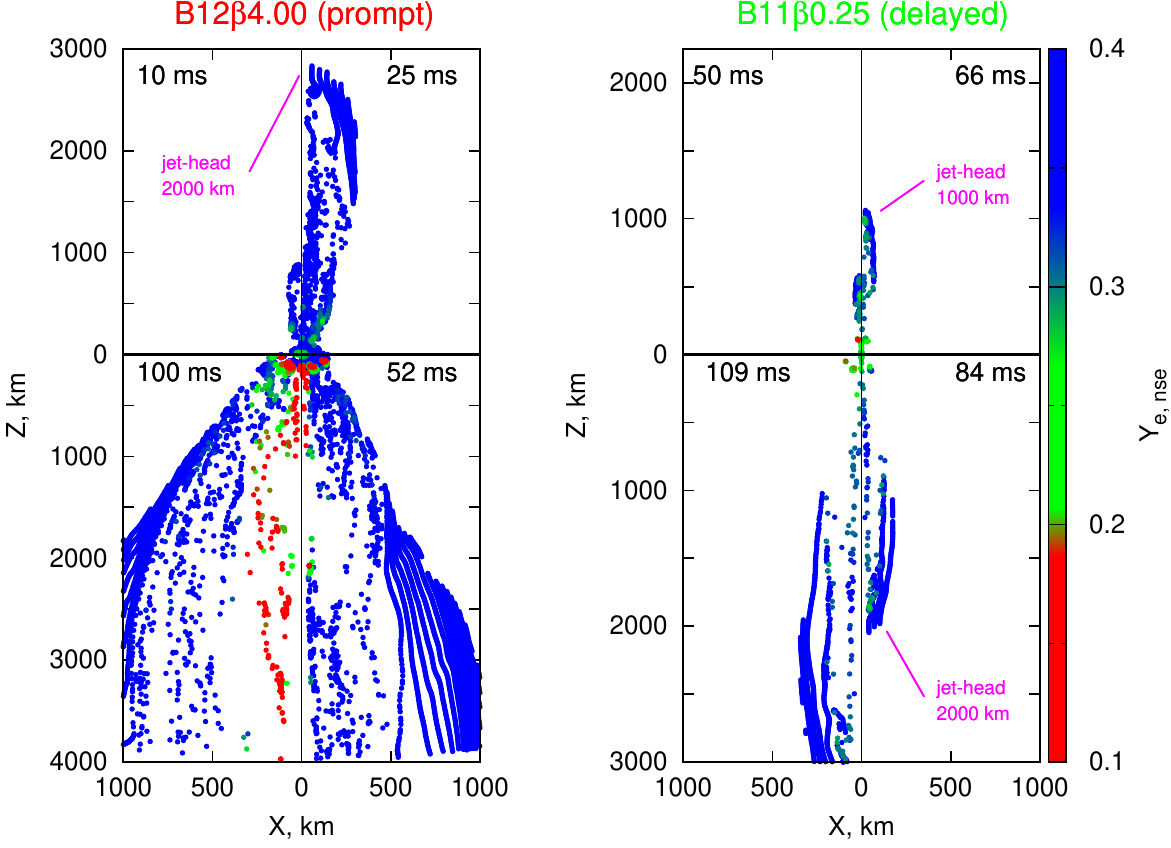}
		\caption{
		The dynamics of explosion models expressed by tracer particles
		for {\tt B12$\beta$4.00} (left) and {\tt B11$\beta$0.25} (right).
		The distribution of particles is shown in a quarter of each panel
		at the moment the jet-head reaches $1000$, $2000$, $3000$, and $4000$~km, respectively.
		Corresponding times after the bounce are shown in each panel:
		which are $10$, $25$, $52$, and $100$~ms for prompt-jets
		and $50$, $66$, $84$, and $109$~ms for delayed-jets, respectively.
		The color scale of each particles indicates $Y_{\rm e,\, nse}$,
		$Y_{\rm e}$ at the end of NSE.}
	\label{fig-jet-evol}
	\end{center}
\end{figure*}

%%%%% Fig. 5 Jet evolution by tracer particles
\begin{figure*}[tbp]
	\begin{center}
		\includegraphics[width=0.775\hsize]{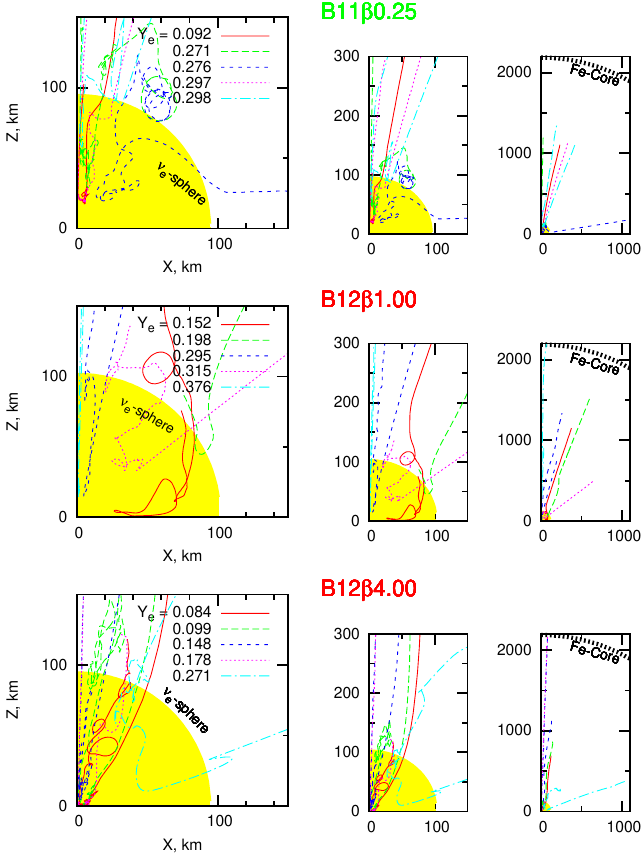}
		\caption{
		The track for ejected tracer particles and the path of tracer particles for a delayed-jet model.
		{\tt B11$\beta$0.25} (upper)
		and two prompt-models, {\tt B12$\beta$1.00} (medium) and {\tt B12$\beta$4.00} (lower),
		are illustrated in different spacial scales.
		Each particle is labeled by $Y_{\rm e}$ at the end of NSE.
		The central filled (yellow) circular area describes the neutrino-sphere.}
	\label{fig-traj}
	\end{center}
\end{figure*}

%.............................................................................................
%        Section 3.2
%.............................................................................................
\subsection{Explosion Models}
\label{sec-expl-model}

Explosion models of the MR-SN are simulated employing the grid-based Eulerian
code described in Section~\ref{sec-srmhd}.
On the other hand, the particle-based Lagrangian hydrodynamic evolution based on the post-process scheme
is utilized (see, Section~\ref{sec-tpm})
to follow the abundance evolution of the models.
Several key physical quantities of the explosion models
are summarized in Table~\ref{tab-expl-properties},
of which names correspond to the pre-collapse models described in Section~\ref{sec-precol}.
The physical properties of dynamics are basically based on the original hydrodynamics models,
although we adopted the TPM for estimating ejecta.

The duration of collapse, $t_{\rm bnc}$, is a time interval in ms
from the beginning of collapse to the core-bounce.
The other three timescales are measured from $t_{\rm bnc}$:
$t_{\rm del}$ is the launch of the MHD jet (delay time),
$t_{\rm exp}$ is the jet reaching the distance of $1000$~km from the PNS,
and $t_{\rm fin}$ is the end of simulation.
Focusing on the outward jet, the velocity is $v_{\rm exp}$
and the corresponding explosion energy is $E_{\rm exp}$ at the time $t=t_{\rm exp}$.
The total amount of ejected matter, estimated by the post-process TPM, is denoted by $M_{\rm ej}$.
There is an alternative mass, denoted by $M_{{\rm ej},\,r}$,
which is the total mass of neutron-rich ejecta (producing heavy $r$-process elements)
for $Y_{\rm e} \leq 0.3$.
Finally, the average electron fraction only for these very neutron-rich ejecta,
$\langle Y_{\rm e} \rangle$,
is defined by
\begin{equation}
	\langle Y_{{\rm e},\,r} \rangle
		= \frac{1}{M_{{\rm ej},\,r}}
			\sum_{i=1}^{\rm all} m_i \, Y_{{\rm e},\,i} \ \ \ ({\rm for }\ Y_{\rm e} \leq 0.3)
\end{equation}
where $m_i$ and $Y_{\rm e, \,\it{i}}$ are the mass and electron fraction (at the end of NSE)
of individual tracer particles, respectively.
This is an index expressing the strength of the $r$-process
that lower values produce heavier nuclei,
mainly because the matter of $Y_{\rm e} \leq 0.3$ contributes to 
the production of heavy $r$-process elements in low entropy conditions of MR-SNe
\citep[see, e.g.,][]{2006ApJ...642..410N, 2008ApJ...680.1350F, 2012ApJ...758....9N, 2012ApJ...750L..22W}.

As described in the previous section, prompt-magnetic-jets, in principle,
have a shorter dynamical timescale, expressed by $t_{\rm del}$ and $t_{\rm exp}$,
with larger $v_{\rm jet}$ and $E_{\rm exp}$
than delayed-jets \citep[for details, see,][]{2009ApJ...691.1360T}.
Additionally, some physical quantities in Table~\ref{tab-expl-properties}
provide a clue to understanding not only the dynamics of the explosion,
but also resulting $r$-process nucleosynthesis.
Focusing on $t_{\rm del}$, the delayed-jet model has a larger value
than prompt-jet models until the launch of the shock.
Here, we can predict a stronger progress for the $r$-process for prompt-magnetic models
than the delayed-jet model by comparing with $Y_{{\rm e}, r}$,
although we need precise network calculations to determine
the detailed nucleosynthesis yields.

%%%%% Fig. 6
\begin{figure}[t]
	\begin{center}
		\includegraphics[width=\hsize]{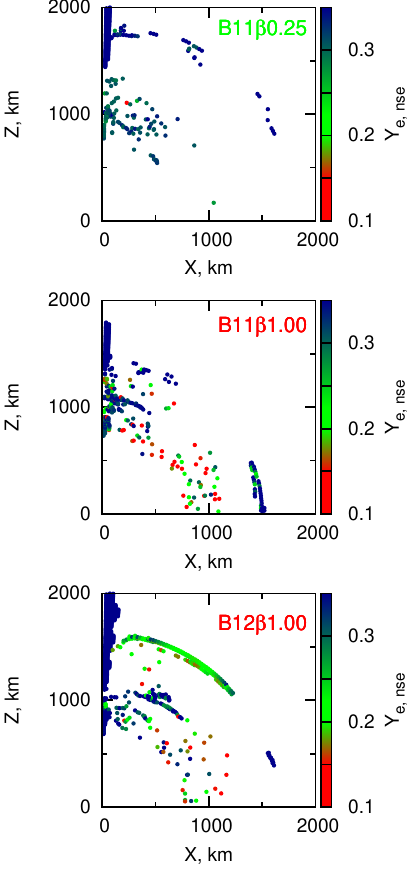}
		\caption{
		The initial position of tracer particles on the meridian plane.
		Ejected particles are plotted with $Y_{\rm e,\ nse}$ at the end of NSE.
		We adopt one delayed-jet ({\tt B11$\beta$0.25})
		and two prompt-jets ({\tt B11$\beta$1.00} and {\tt B12$\beta$1.00}).}
	\label{fig-ini-map}
	\end{center}
\end{figure}

\subsubsection{Dynamics of Tracer Particles}

The time evolution of tracer particles for the ejected matter is shown
for the both of the prompt-jet and delayed-jet in Figure~\ref{fig-jet-evol}.
The $Y_{\rm e,\,nse}$, which indicates $Y_{\rm e}$ at the end of NSE,
is also shown for each particle.
Each panel shows the evolution of the explosion model in different times,
for {\tt B12$\beta$4.00} (left) and {\tt B11$\beta$0.25} (right),
representing the prompt-jet and delayed-jet, respectively.
A quarter of each panel corresponds to the distribution of tracer particles at a given time,
in the time sequence, which goes through upper left, upper right,
lower right, and lower left, respectively, evolved in clockwise.
The time after the bounce is shown in each panel:
$10$, $25$, $52$, and $100$~ms for the prompt-jet
and $50$, $66$, $84$, and $109$~ms for the delayed-jet, respectively.
The value of $Y_{\rm e}$ distributes
in the range of $0.1$--$0.4$ and $0.2$--$0.4$
for the prompt model and delayed model, respectively.

As clearly seen in Figure~\ref{fig-jet-evol},
the prompt-jet tends to have a broader structure,
of which the jet-head reaches $\sim 3000$~km at $25$~ms after the bounce,
whereas the collimated jet of the delayed model takes $\sim 100$~ms to reach $3000$~km.
Focusing on the $Y_{\rm e}$, we can see that neutron-rich matter (i.e., $Y_{\rm e} < 0.3$),
is ejected in jets along the rotational axis for all the models.
Neutron-rich tracer particles in the prompt-jet model
are expected in the body of the jet-like outflow,
and are continuously launched following the initially ejected matter.
The delayed model has a different feature:
which moderate neutron-rich particles are mostly populated in a jet-head region,
and are ejected by the primary jet-like outflow.
By only focusing on bulk motion of tracer particles, however,
we cannot fully understand the ejection mechanism of neutron-rich matter.

%%%%% Fig. 7
\begin{figure*}[tbp]
	\begin{center}
		\includegraphics[width=\hsize]{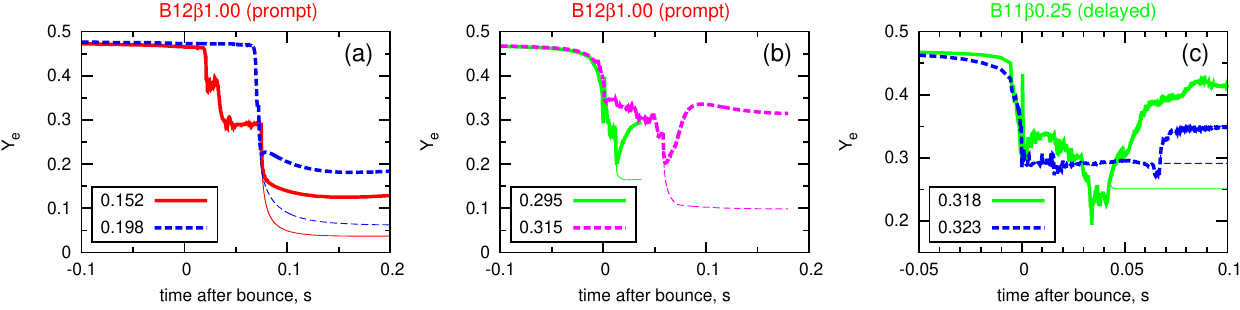}
		\caption{
		Time evolution of the $Y_{\rm e}$ for the selected tracer particles
		from {\tt B12$\beta$1.00} (prompt-jet) and {\tt B11$\beta$0.25} (delayed-jet).		
		Each line begins at the onset of collapse,
		of which time is measured from at the core-bounce.
		Thick lines are results taking into account all weak reactions
		between the nucleons and neutrinos,
		whereas thin lines are results ignoring neutrino absorptions.}
	\label{fig-ye-evol}
	\end{center}
\end{figure*}

%%%%% Fig. 8
\begin{figure*}[tbp]
	\begin{center}
		\includegraphics[width=0.75\hsize]{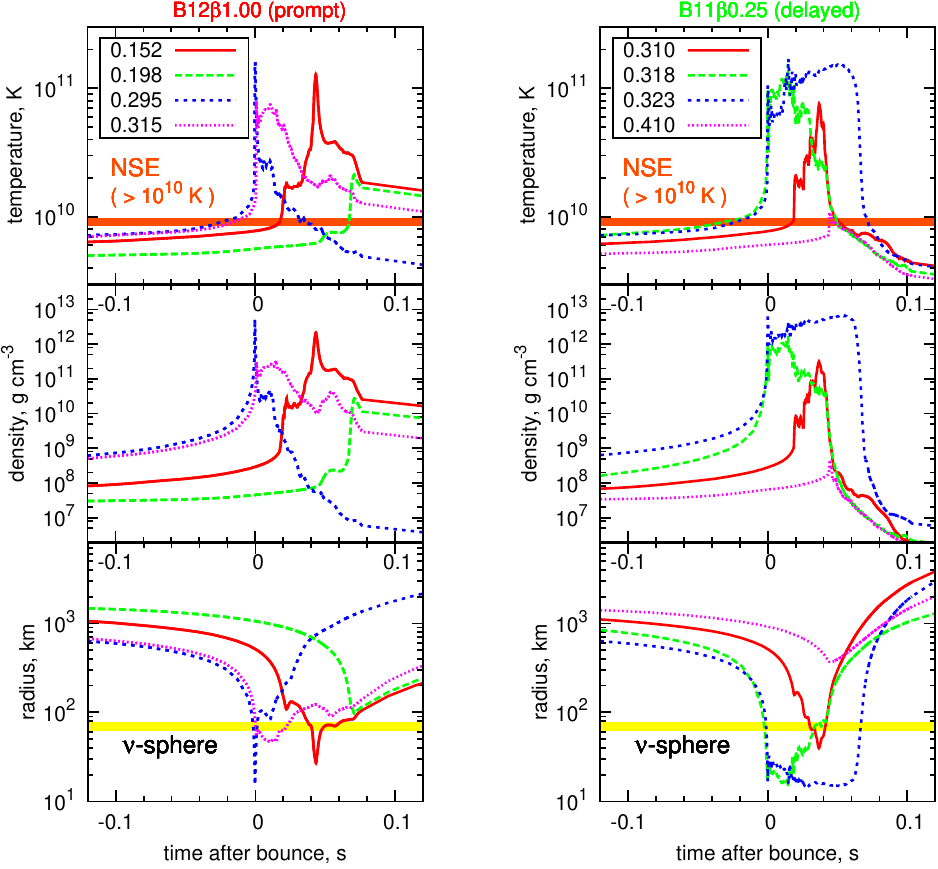}
		\caption{
		Time evolution of hydrodynamical values
		for {\tt B12$\beta$1.00} (prompt; left) and {\tt B11$\beta$0.25} (delayed; right).
		Each panel includes the temperature in K (top),
		density in g ${\rm cm}^{-3}$ (medium),
		and radius in km (low) for selected tracer particles.
		The condition for the NSE ($T \sim 9$~GK)
		and the boundary of the neutrino-sphere are highlighted.}
	\label{fig-hydro-evol}
	\end{center}
\end{figure*}

Therefore, we describe detailed ejection processes
using selected tracer particles with a path on the meridian ($X$--$Z$)
plane, as illustrated in Figure~\ref{fig-traj}.
Three different explosion models are adopted:
the delayed-jet model, {\tt B11$\beta$0.25}, and two prompt-jets,
{\tt B12$\beta$1.00} and {\tt B12$\beta$4.00}.
Each explosion model has three panels in different spacial scales,
and a central area with a radius of $\sim 100$~km corresponding to the surface of the neutrino-sphere.
Each figure has paths of six selected tracer particles that follow the history of movement
labeled by $Y_{\rm e, nse}$, i.e.,
which is the constant of $Y_{\rm e}$ at the end of NSE ($T=9$~GK).
All presented trajectories, in principle, fall toward the center during core-collapse
and escape along the rotational axis while experiencing
chaotic convectional motion around the central core.

The type of prompt-jets represented by
{\tt B12$\beta$1.00} and {\tt B12$\beta$4.00}
has an energetic jet-like outflow with a wider opening angle,
as seen in Figure~\ref{fig-jet-evol}.
Tracer particles from these explosion models may have
low $Y_{\rm e, nse}$ along the rotational ($Z$) axis.
As shown in Figure~\ref{fig-traj}, the most neutron-rich ejecta in a range of
$Y_{\rm e, nse} = 0.1$ -- $0.2$ escape after they deeply fall into the PNS.
For the case of {\tt B12$\beta$4.00}, which is the most energetic model,
a large amount of particles are ejected from the central region ($\sim 10$~km) inside the PNS.
By the comparison with {\tt B12$\beta$1.00} and {\tt B12$\beta$4.00},
neutron-rich ejecta tend to move the central region, which requires a higher explosion energy to eject them.
On the other hand, moderate neutron-rich ($Y_{\rm e, nse} = 0.2$--$0.3$) ejecta
are predominant in the delayed-magnetic-jet explosion ({\tt B11$\beta$0.25}).
For the delayed model, trajectories fall into around the PNS core
and then escape from the core along the rotational axis
after a long duration ($\sim 50$~ms) of convective motion.
This duration corresponds to the enhancement process of magnetic fields
by field wrapping due to strong differential rotation \citep{2009ApJ...691.1360T},
which results in a higher magnetic pressure to overcome the ram pressure of falling matter.

All ejected particles of our explosion models finally escape along the rotational axis,
although the convective motion in the central area is quite complicated.
In order to see the initial position of neutron-rich ejecta,
we illustrate the distribution of tracer particles at the beginning of collapse
with $Y_{\rm e, nse}$ in Figure~\ref{fig-ini-map}.
Similar to Figure~\ref{fig-traj},
we selected {\tt B11$\beta$0.25}, {\tt B11$\beta$1.00},
and {\tt B12$\beta$1.00}, representing the delayed-magnetic jets
and prompt-magnetic jets, respectively.
The radius of the iron core for the progenitor model is $\sim 2000$~km.
Thus, this figure shows that ejected particles, in which $Y_{\rm e, nse}$ is significantly below $0.5$,
are initiated in the iron core.
Particles initially located outside the core along the rotational axis are ejected,
avoiding the progress of weak interactions (i.e., electron-captures),
which remain their initial $Y_{\rm e}$ (i.e., $Y_{\rm e, nse}\sim 0.5$).
For the delayed model, the lowest $Y_{\rm e, nse} \sim 0.3$,
which was initially located $1000$~km from the center,
spreads in the direction of $45^{\circ}$ from the axis.
By contrast, prompt-jets have an initial distribution
that from the axis to the equatorial direction,
especially neutron-rich ejecta that is $Y_{\rm e, nse} \sim 0.1$,
which initially locates not only the $Z$-direction, but also the $X$-direction near the equatorial plane.
For all of the models, tracer particles initiated within the $500$~km of the iron core
immediately fall into the center and finally construct a PNS,
which has never been ejected.

%%%%% Fig. 9
\begin{figure}[tbp]
	\begin{center}
		\includegraphics[width=\hsize]{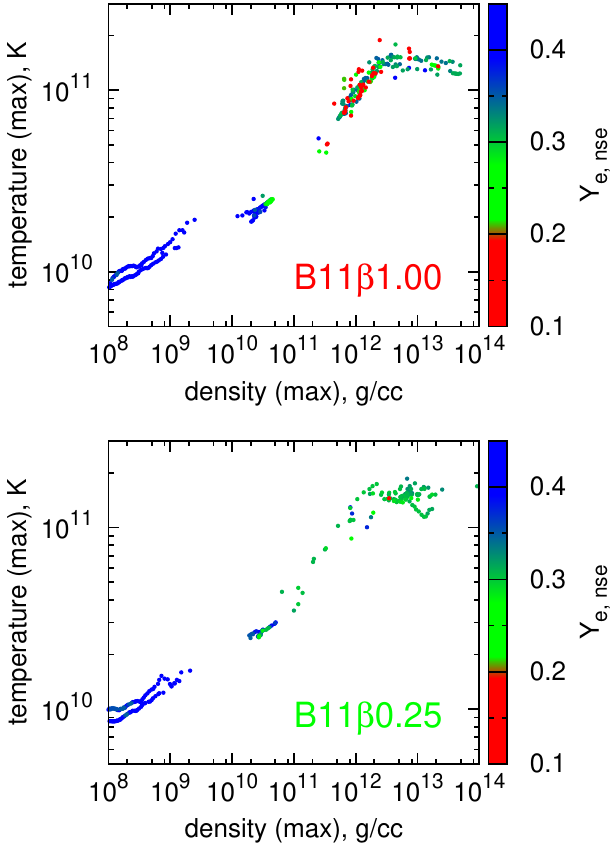}
		\caption{
		The maximum density and temperature of ejecta with $Y_{\rm e}$
		for a prompt model, {\tt B11$\beta$1.00} (upper),
		and a delayed model, {\tt B11$\beta$0.25} (lower).
		The $Y_{\rm e, nse}$ of each particle is shown in the given color range.}
	\label{fig-max-phys}
	\end{center}
\end{figure}

%%%%% Fig. 10
\begin{figure*}[htbp]
  \begin{center}
    \includegraphics[width=0.89\hsize]{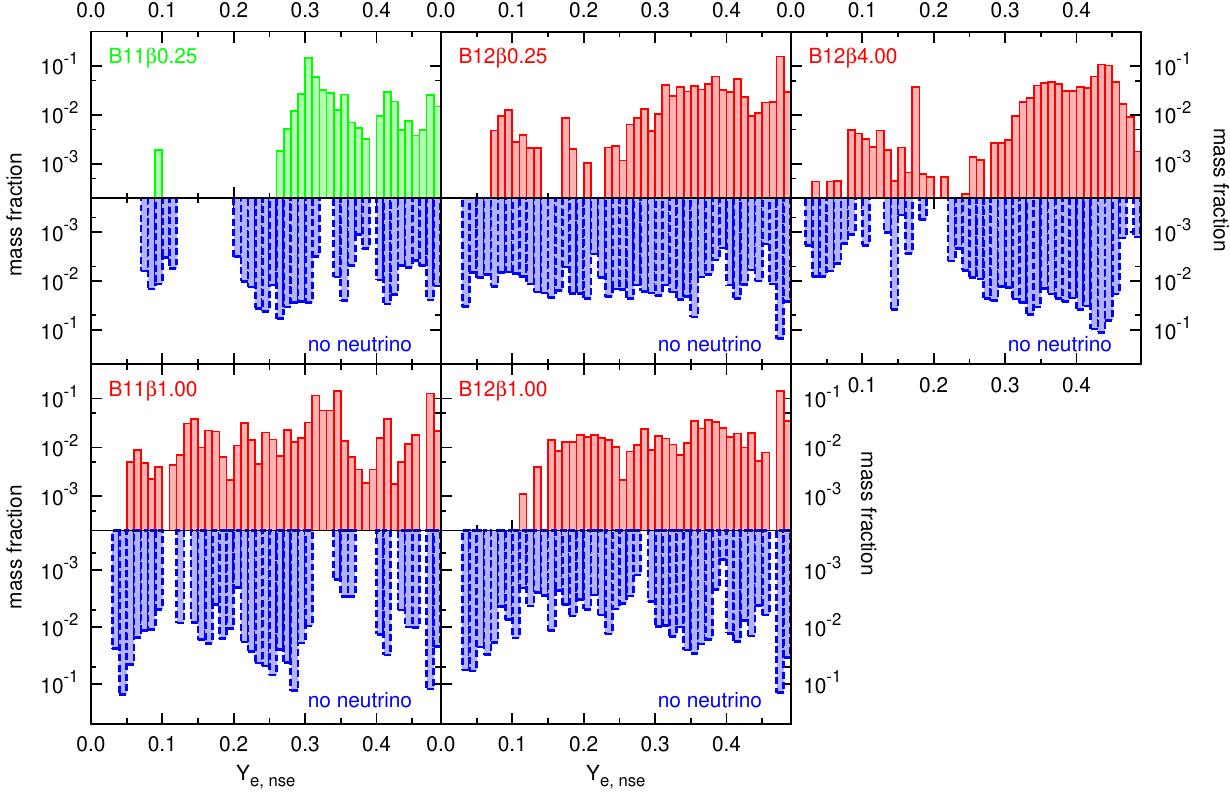}
    \caption{
      The histogram of $Y_{\rm e,nse}$ vs. the mass fraction of the total ejecta.
      Each plot corresponds to the explosion model whose name is labeled in the upper left,
      where there are two different histograms, that is,
      the case includes neutrino absorptions (red bars) and ignores these effects (blue bars).}
    \label{fig-ye-hist}
  \end{center}
\end{figure*}

%%%%% Fig. 11
\begin{figure*}[htbp]
  \begin{center}
    \includegraphics[width=0.89\hsize]{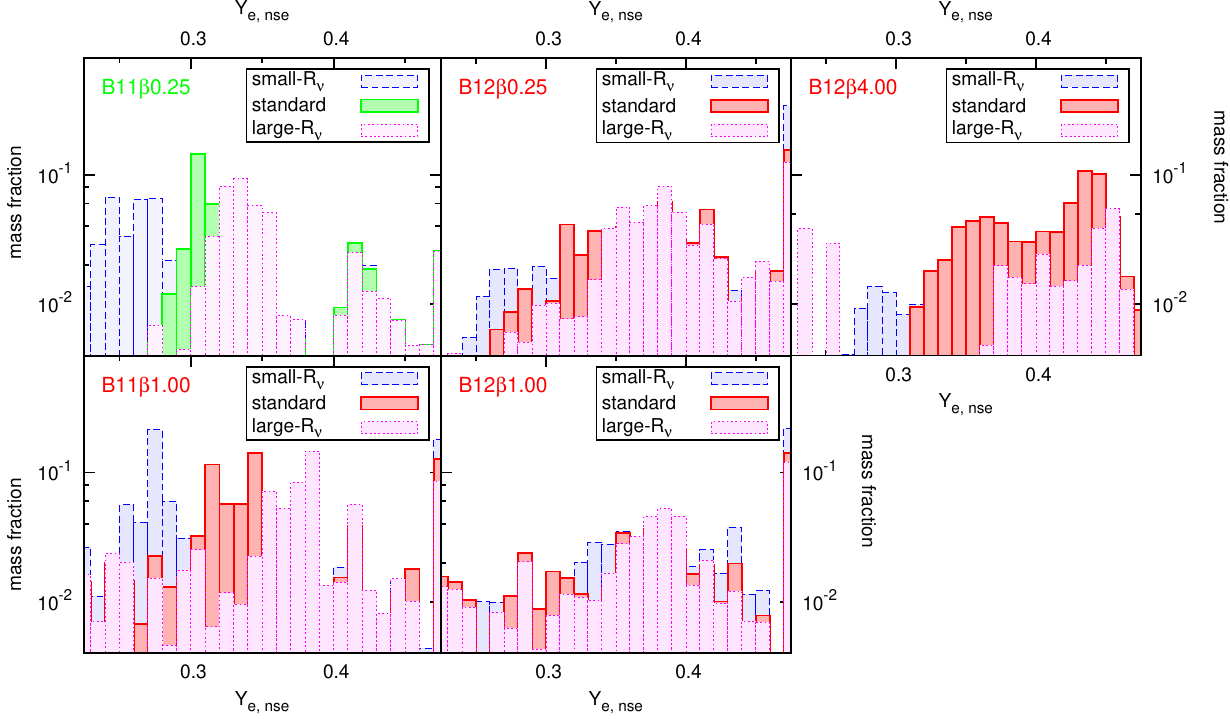}
    \caption{The histogram of $Y_{\rm e}$ for all ejecta changing the radius of the neutrino-sphere
      where the mass fraction is plotted in linear scale.
      Each panel has two bars in addition to those in Figure~\ref{fig-ye-hist}
      (green for the delayed model and red for prompt models),
      for which we assume a longer (blue) and shorter (magenta)
      time duration of neutrino absorptions, respectively.}
    \label{fig-ye-uncertainty}
  \end{center}
\end{figure*}

\subsubsection{Ejection Process of Neutron-rich Matter}
\label{sec-traj-evol}

We examine the ejection process of neutron-rich matter in detail,
which corresponds to the evolution of $Y_{\rm e}$,
focusing on representative ejecta (tracer particles).
The time evolution of the $Y_{\rm e}$ for the selected tracer particles
is shown in Figure~\ref{fig-ye-evol},
based on {\tt B12$\beta$1.00} (prompt) and {\tt B11$\beta$0.25} (delayed), respectively.
All panels have the evolution history of $Y_{\rm e}$ for the two tracer particles
labeled by $Y_{\rm e, nse}$ and computed by TPM.
Each particle has two different lines,
where the thick line shows evolution of the $Y_{\rm e}$
including all weak interactions described in Section~\ref{sec-tpm-ye},
and the thin line indicates the result of the case ignoring neutrino absorption.
We plot the cases of prompt-jet explosions in (a) and (b) of Figure~\ref{fig-ye-evol}.
These tracer particles correspond to very neutron-rich matter (a), $Y_{\rm e} < 0.2$,
as well as mild neutron-rich (b), $Y_{\rm e} \sim 0.3$, respectively.
The evolution of $Y_{\rm e}$ based on the delayed explosion model
is also plotted in (c), where the final $Y_{\rm e} > 0.3$.

We can qualitatively understand the role of neutrino absorption on the evolution of $Y_{\rm e}$.
For all the cases, the $Y_{\rm e}$ of the ejecta rises via the neutrino absorption,
where the reaction ${\rm n} + \nu_{\rm e} \rightarrow {\rm p} + {\rm e}^- $
is predominant for neutron-rich ($Y_{\rm e} < 0.5$).
The activity of this reaction is determined by the distance from the PNS (neutrino-sphere),
and the luminosity and mean energy of the emitted neutrino,
which appear to be independent of the geometry of explosion (the direction of ejection).
Thus, during the expansion phase, the effect of neutrino absorption is well explained,
assuming the spherical symmetry geometry,
where reaction rates are just determined by the distance,
even in cases of multi-dimensional hydrodynamics.
Each tracer particle increases the $Y_{\rm e}$ (thick line) due to neutrino absorption
up to three times compared with the case without neutrino absorption (thin line).
The physical uncertainty of neutrino reactions due to the properties of PNS
is discussed in the following section.

The neutron-richness, which is in inverse proportion to the $Y_{\rm e}$,
is determined by the strength of the electron captures on protons,
which become stronger in higher densities.
In order to examine more details,
we now describe the hydrodynamical evolution of the temperature,
density, and radial distance from the center
for the tracer particles shown in Figure~\ref{fig-hydro-evol}.
Neutron-rich ejecta in the prompt-jet model,
$Y_{\rm e,\,nse} = 0.152$ and $0.198$
in Figure~\ref{fig-ye-evol}a and left panel of Figure~\ref{fig-hydro-evol},
have higher maximum densities ($> 10^{12}~\rm{g}~{\rm cm}^3$) at the bounce,\footnote{
In this context, we use the term ``bounce'' for each particle,
which does not correspond to the time of bounce at the PNS core.}
where the electron capture is more active than other weak interactions.
On the other hand, the particles, $Y_{\rm e,\,nse} = 0.295$ and $0.315$ in Figure~\ref{fig-ye-evol}b,
have lower densities in the early stage of the expansion.
These $Y_{\rm e}$s increase after the bounce,
which implies that the effect of neutrino absorption is stronger than electron captures.

The behavior of the $Y_{\rm e}$ evolution for the delayed model appears to be rather complicated.
The ejecta initially stays in the PNS core, which has a higher density ($\sim 10^{12}$~g~${\rm cm}^{-3}$)
and temperature ($\sim 10^{11}$~K),
whose $Y_{\rm e}$ is determined by the $\beta$-equilibrium condition.
These particles remain in the core until $0.05$ s ($= 50$~ms) after the (failed) bounce,
where this time duration corresponds to the timescale of the magnetic fields enhancement ($t_{\rm del}$)
in Table~\ref{tab-expl-properties}.
The density and temperature immediately drop, resulting in the lower densities,
during the later phase of the explosion.
The $Y_{\rm e}$ sharply increases at the beginning of the expansion,
due to strong neutrino absorption, because they are close to the neutrino-sphere
\citep[also found by][]{2012ApJ...750L..22W}.
Here, we should note that the effects of neutrino absorption (weak interaction)
are sensitive to the treatment of neutrino reactions and the transport,
especially for the delayed-jet models.
More sophisticated treatments are desirable for more precise predictions
for the $Y_{\rm e}$ of ejecta.

The maximum density and temperature of the tracer particles with the $Y_{\rm e,\,nse}$
are plotted in Figure~\ref{fig-max-phys},
for {\tt B11$\beta$1.00} (prompt) and {\tt B11$\beta$0.25} (delayed), respectively.
The time of the maximum density almost corresponds
the time of the maximum temperature,
which is also close to the bounce time of each particle.
In each panel of Figure~\ref{fig-max-phys},
particles in higher densities and temperatures (upper right)
correspond to the ejecta that has escaped from deepest regions inside the PNS,
of which the matter finally remains neutron-rich ($Y_{\rm e, nse} < 0.3$).
A group of particles in the low densities and temperatures (lower left),
on the other hand, is composed of ejecta in the jet-cocoon
pushed by the jet-like outflow without collapse.
The latter component of the ejecta remains initial $Y_{\rm e}$ $( > 0.4)$
at the pre-collapse stage, where the $r$-process hardly occurs.

The values of $Y_{\rm e}$ for neutron-rich matter are different between the prompt and delayed models,
although they have a similar distribution of the maximum density and temperature (Figure~\ref{fig-max-phys}).
As we have described, prompt-jets have very neutron-rich ejecta of $Y_{\rm e, nse} < 0.2$,
whereas the ejecta of the delayed-jet hardly falls below $Y_{\rm e, nse} \sim 0.3$.
This is caused by the difference of the ejection processes after the core-bounce,
which is independent from the maximum values of densities and temperatures.
Here, we recall that the timescale for the amplification of magnetic fields
($t_{\rm del}$ in Table~\ref{tab-expl-properties}, or the duration of magnetic fields wrapping)
is a crucial factor to determine the property of the explosion models around the PNS.
Additionally, the delayed model has a lower explosion energy,
in which the densities and temperatures are lower,
leading to weaker electron captures on the protons.
The delayed model in the present study, which is based on a long-term SR-MHD simulation,
has been overlooked in previous nucleosynthesis studies
because they mostly focused on strong magnetic driven jets
corresponding to prompt-jets in the present classification.
However, delayed-jet models, which are expected to have different elemental products,
cannot be ignored for the nucleosynthesis study of MR-SNe.
Finally, we should note that the property of the PNS is sensitive to relevant physical processes.
We evaluate the uncertainty of $Y_{\rm e}$ based on our current explosion models in the following section.

%%%%% Fig. 12
\begin{figure*}[htbp]
	\begin{center}
		\includegraphics[width=1.0\hsize]{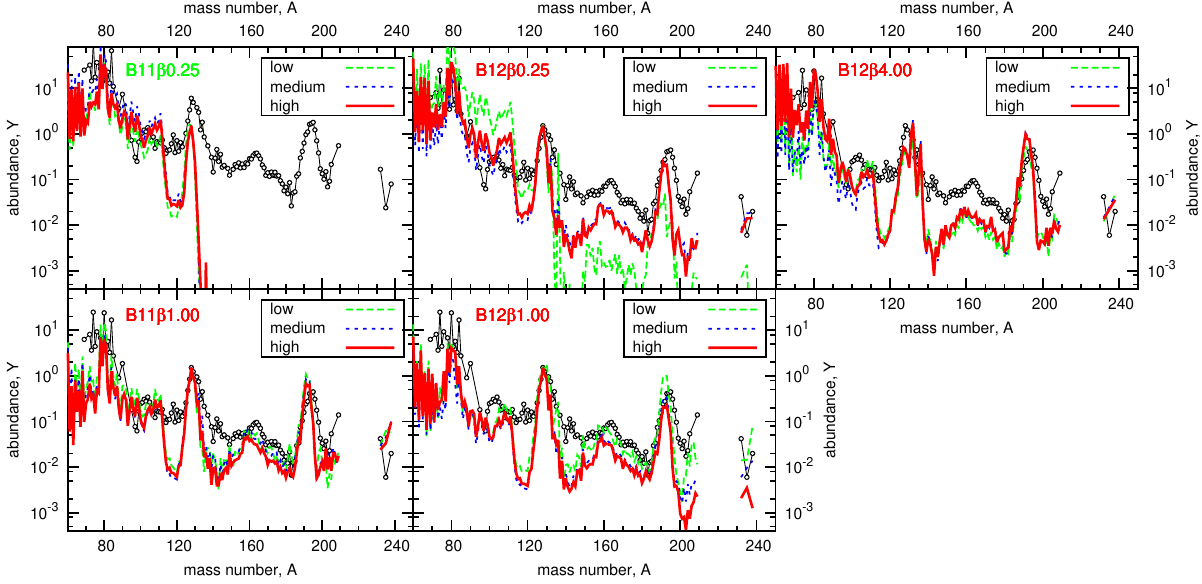}
		\caption{The final abundances ($Y$) as a function of mass number ($A$)
		with the solar $r$-process pattern (open dots).
		Each explosion model has three different lines,
		which are the results of low (green), medium (blue), and high (red)
		resolution for adopted tracer particles.}
	\label{fig-final-abund}
	\end{center}
\end{figure*}

\subsubsection{The $Y_{\rm e}$ of ejecta and uncertainty of PNSs}
\label{sec-ye}

We have calculated the evolution of $Y_{\rm e}$ using the TPM for all our explosion models.
Figure~\ref{fig-ye-hist} shows the histograms of the $Y_{\rm e}$ (at the end of NSE)
for the entire ejecta in a range of $0 < Y_{\rm e,\; nse} < 0.45$,
where the width of $Y_{\rm e}$ ($Y_{\rm e}$-bin) corresponds to $\Delta Y_{\rm e}= 0.02$.
Each model has two different $Y_{\rm e}$-histograms,
which take into account all relevant weak reactions
(green or red bars of the upper panel) and ignore neutrino absorption
during expansion (blue bars of the lower panel), respectively.

Prompt-jets (red histograms in Figure~\ref{fig-ye-hist}) have lower $Y_{\rm e}$ ejecta
compared with the delayed model (green), of which peak at $Y_{\rm e ,nse} \sim 0.3$.
The prompt-jet has a variety of $Y_{\rm e}$ distribution among all explosion models.
We can confirm that all prompt models have very neutron-rich ejecta of $Y_{\rm e, nse} < 0.2$,
and expect successful $r$-processes producing the third peak.
As we show in the following nucleosynthesis calculations,
the difference in the $Y_{\rm e, nse}$ among all prompt-jet models becomes small
when we see the resulting nucleosynthesis abundance patterns.
On the other hand, the delayed model (the green histogram in Figure~\ref{fig-ye-hist})
mostly has moderate neutron-rich ejecta $Y_{\rm e, nse} \sim 0.3$,
of which the production of heavy $r$-process elements is restricted.
This difference of the $Y_{\rm e}$-histogram between the prompt and delayed models
remains significant in the final abundance pattern.

Here, we will emphasise the importance of neutrino absorption in MR-SN models
for determining of $Y_{\rm e}$.
In Figure~\ref{fig-ye-hist}, the histogram in the lower panel of each model shows result
that we ignored the effect of neutrino absorption during expansion (the neutrino-optically thin region).
The $Y_{\rm e}$ generally increases due to neutrino absorption in realistic explosion conditions,
in which we have shown the time evolution of selected tracer particles in Figure~\ref{fig-ye-evol}.
By comparing these two $Y_{\rm e}$-histograms,
we find that the peak of extremely low $Y_{\rm e} < 0.1$ in the case of ignoring neutrino absorption
is removed or shifted to higher $Y_{\rm e}$
(the difference is $\Delta Y_{\rm e} \sim 0.1$)
when we take into account neutrino absorptions.
Although detailed treatments of weak reactions are essential to calculate $Y_{\rm e}$,
the precise treatments of neutrino transport fully coupled with multi-dimension hydrodynamics
are still computationally expensive.

Therefore, we take into account the physical uncertainties relevant to the weak interaction and the PNS
by changing the radius of the neutrino-sphere when we calculate the evolution of $Y_{\rm e}$ in the TPM.
Based on the standard value, we assume a larger and smaller radius,
which corresponds to the shorter and longer time duration of neutrino absorption,
respectively.
The fiducial values of the radius are generally
$R_{\nu_{\rm e}} \sim 100$, as shown in Table~\ref{tab-neutrino},
where we assume the variation of $\pm 20$\% (i.e., the differences are $\sim 10$~km).

The result of $Y_{\rm e, nse}$, based on the the uncertainty of the PNS,
is shown in Figure~\ref{fig-ye-uncertainty},
which plots the histogram of $Y_{\rm e}$ for all explosion models.
They generally exhibit a lower $Y_{\rm e}$ distribution for a smaller radius case,
whereas a larger radius of the neutrino-sphere results in higher values for the $Y_{\rm e}$.
This is because a larger neutrino-sphere causes early ejection from the neutrino-sphere
(Phase~1 in Figure~\ref{fig-n-sphere}),
which means electron captures reducing the $Y_{\rm e}$ are active.
The variation of the neutrino-sphere radius also affects the reaction rates
of the neutrino interaction outside the neutrino-sphere,
in which the smaller radius results in weaker neutrino absorption.
In contrast, the case of the larger neutrino-sphere radius
has the longer duration staying the neutrino-sphere
(i.e., tracer particles are influenced by weaker electron captures and stronger neutrino absorption).
The delayed model is basically affected by neutrino absorption,
so that the effect of uncertainty on the $Y_{\rm e}$ is larger than prompt-jets,
as shown in Figure~\ref{fig-ye-uncertainty}.

% -----------------------------------------------------------------------------------------------
\section{Nucleosynthesis}
\label{sec-res}

Nucleosynthesis calculations are performed based
on the MR-SN models described in the previous section.
The results are compared with several $r$-process abundance patterns
of the sun and metal-poor stars.
We discuss the effects of physical uncertainties on nucleosynthesis,
focusing on nuclear reactions, properties of PNS, and hydrodynamical instabilities of jets.
Finally, the production of $^{56}$Ni in jet-like explosions of the MR-SNe is briefly discussed
with optical observation of SNe.

%................................................................................................
\subsection{The $r$-process Nucleosynthesis}

We performed nucleosynthesis calculations employing the nuclear reaction network code
described in Section~\ref{sec-network}.
The network calculations are carried out by following the thermodynamical evolution
of trajectories in the basis of the MR-SNe models.
All of results for the explosion models are shown in Figure~\ref{fig-final-abund},
which are plotted integrated final abundances $Y$
(the number abundance relative to the Si atoms, $Y(\rm Si) = 10^6$ )
as a function of mass number $A$ of each isobar.
Each panel has three lines for different resolution,
which are labeled ``low,'' ``medium,'' and ``high,''
corresponding to the sets of tracer particles in Table~\ref{tpm-init}.
Observational-based solar $r$-process abundances
\citep[the residual of the classical s-process by][]{1999ApJ...525..886A}
are depicted by open dots.

We confirm that the nucleosynthesis calculations are numerically converged
because the abundances of high and intermediate (medium) resolution are almost identical.
In the case of low resolution, we can see that some models are not converged.
The models of {\tt B12$\beta$0.25} and {\tt B12$\beta$1.00}, in particular,
have a significant difference in abundance patterns between the low resolution and higher resolution calculations.
In general, the number of tracer particles necessary to resolve the final
$r$-process abundances depends on the hydrodynamical properties
of explosion models and the treatment of tracer particle motion.
Our results point out that several hundred particles (the current medium resolution)
in 2D hydrodynamics simulations are required to resolve the resulting $r$-process,
which has not been satisfied in previous studies.

The mass of $r$-process elements in the ejecta and the total ejected masses of a jet
are summarized in Table~\ref{tab-rproc} for all explosion models.
The total mass of the ejecta is estimated only for the jet-like outflow,
which is located in $\leq 4000$~km at the pre-collapse phase.
The total mass of the $r$-process material is in the range of $0.957 \times 10^{-2}$ -- $2.45 \times 10^{-2} M_\odot$,
where the less energetic delayed model of {\tt B11$\beta$0.25} shows a relatively lower value
and the most energetic prompt model, {\tt B12$\beta$4.00}, has a much larger value.
These higher masses are necessary for ejecting very neutron-rich matter
from the deeper region of a PNS, avoiding the $Y_{\rm e}$ equilibrium,
which is different from the moderate neutron-rich ejecta of PNS winds.
Using an analytic formula of the minimum mass loss rate for a neutron-rich ejecta
\citep[Equation~19 of][]{2008ApJ...676.1130M},
we can derive $\sim 1 \times 10^{-2} M_\odot~{\rm s}^{-1}$ adopting typical values of Table~\ref{tab-rproc},
where $L_{\nu_{\rm e}} \sim 4 \times 10^{52}~{\rm erg}~{\rm s}^{-1}$,
$R_{\nu_{\rm e}} \sim 80~{\rm km}$, and $E_{\nu} \sim 15~{\rm MeV}$
with the $1.4 M_\odot$ PNS mass.
This mass loss rate results in $\sim 1 \times 10^{-2} M_\odot$,
with the duration of mass ejection of $200$ ms,
which is consistent with results from the hydrodynamics simulations.

%%% table 5
\begin{table}[tbp]
\centering
\caption{Masses of $r$-process Elements and the Total Ejecta}
\begin{center}
\begin{tabular*}{\hsize}{@{\extracolsep{\fill}}ccc}
\hline
\hline
Model
& $r$-process Mass & Ejected Mass \\
& ($10^{-2} M_{\odot}$) & ($10^{-2} M_{\odot}$) \\
\hline
    {\tt B11$\beta$0.25} & $0.957$ & $2.68$ \\
    {\tt B11$\beta$1.00} & $1.53$  & $2.15$ \\
    {\tt B12$\beta$0.25} & $1.26$  & $3.55$ \\
    {\tt B12$\beta$1.00} & $1.66$  & $4.37$ \\
    {\tt B12$\beta$4.00} & $2.45$  & $8.57$ \\
    \hline
\end{tabular*}
\end{center}
\tablecomments
    {The amount of $r$-process elements in the ejecta
      and the total ejected mass in the solar mass for our MR-SN explosion models.}
\label{tab-rproc}
\end{table}

The higher mass ejection of the $r$-process elements
supports the importance of MR-SNe in the galactic chemical evolution.
The typical ejected mass of $r$-process nuclei in MR-SNe, $\sim 10^{-2} M_\odot$,
is several thousand times larger than the typical ejected mass of PNS winds
(the case of canonical CC-SNe).
These higher values of $r$-process elements indicate
that MR-SNe have a significant impact on
the galactic chemical evolution if their event rate is very low.
If we assume that only MR-SNe are the source of $r$-process elements,
we can estimate that $\sim 0.1 \%$ of all CC-SNe explain the amount of the Galactic $r$-process material
by simple multiplication.

%%%%% Fig. 13
\begin{figure}[t]
	\begin{center}
		\includegraphics[width=\hsize]{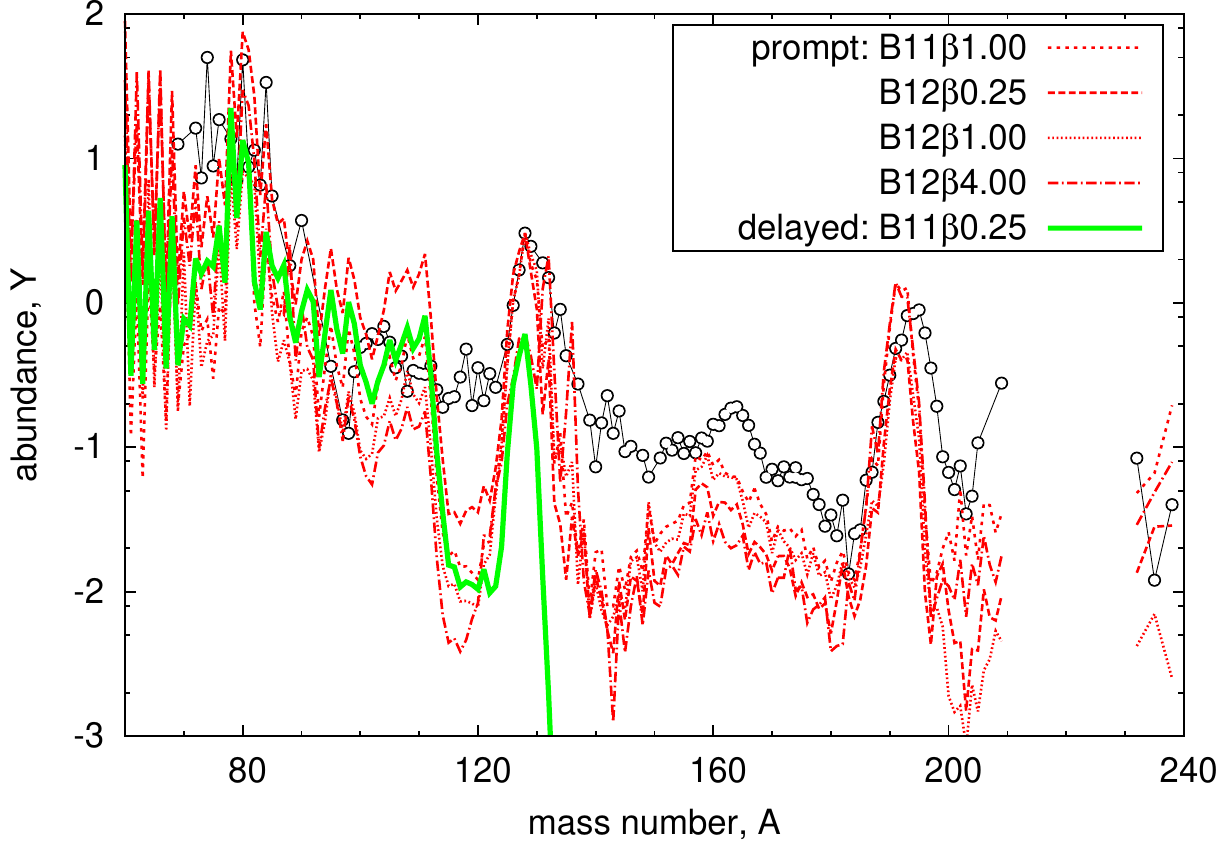}
		\caption{
		The final abundances of all explosion models are plotted with
		the solar $r$-process pattern \citep{1999ApJ...525..886A}.
		Red lines indicate the results for the prompt-jet
		and the green line corresponds to the delayed-jet.}
	\label{fig-final-sol}
	\end{center}
\end{figure}

\subsubsection{The Prompt-jet versus the Delayed-jet}

Comparing the final abundance patterns of all models shown in Figure~\ref{fig-final-sol},
we find common global features in prompt-jets;
these are all models with the exception of {\tt B11$\beta$0.25},
which is the only delayed model.
Prompt-jet models, which are {\tt B11$\beta$1.00},
{\tt B12$\beta$0.25}, {\tt B12$\beta$1.00}, and {\tt B12$\beta$4.00},
successfully produce heavy $r$-process nuclei,
including the second and third peaks and actinides,
although the abundances of lighter isotopes have dispersion.
On the other hand, the final abundance curve of the delayed model ({\tt B11$\beta$0.25})
reaches up to the second $r$-process peak,
where heavier nuclei ($A > 100$) are severely underproduced.
These similarities and the difference in the final abundances among our MR-SN models
are consistent with the discussion in Section~\ref{sec-ye},
which is based on the $Y_{\rm e}$-histograms in Figure~\ref{fig-ye-hist}.

The final abundances of Figure~\ref{fig-final-sol}
are the results of the nucleosynthesis calculations
based on the same nuclear physics input
(i.e., experimental and theoretical masses, reaction rates, and decay properties).
Therefore, each nucleosynthesis abundance pattern reflects the property of each hydrodynamical model,
determined by stellar rotation and magnetic fields.
Our results clearly indicate that the rotation and magnetic fields of stellar models
change the $r$-process nucleosynthesis,
that is, the nucleosynthesis signature of the MR-SNe has a wide variety of observable features (chemical abundances),
caused by the dynamical properties of stellar evolution models.
In particular, the prompt and delayed models appear to be distinguished
in the characteristics of the $r$-process yields.

All of the prompt-jet models, which are energetic jet-like explosions,
have the similar feature of the $r$-process yields,
although the initial conditions and the details of the following explosion process are different.
On the other hand, the weaker explosion of the delayed-jet
produces elements up to the second peak.
The physical properties of MR-SN explosion models continuously vary,
depending on the initial rotation and magnetic fields.
Thus, the transition of nucleosynthesis yields from the prompt-jet to the delayed-jet
appears to indicate the existence of a threshold
for the production of $r$-process nuclei.
Although we found a qualitative difference between the prompt and delayed-jets,
the number of explosion models in the current study is insufficient to identify these thresholds.
Further systematic parameter studies are necessary for the clarification.

%%%%% Fig. 14
\begin{figure*}[htbp]
	\begin{center}
		\includegraphics[width=\hsize]{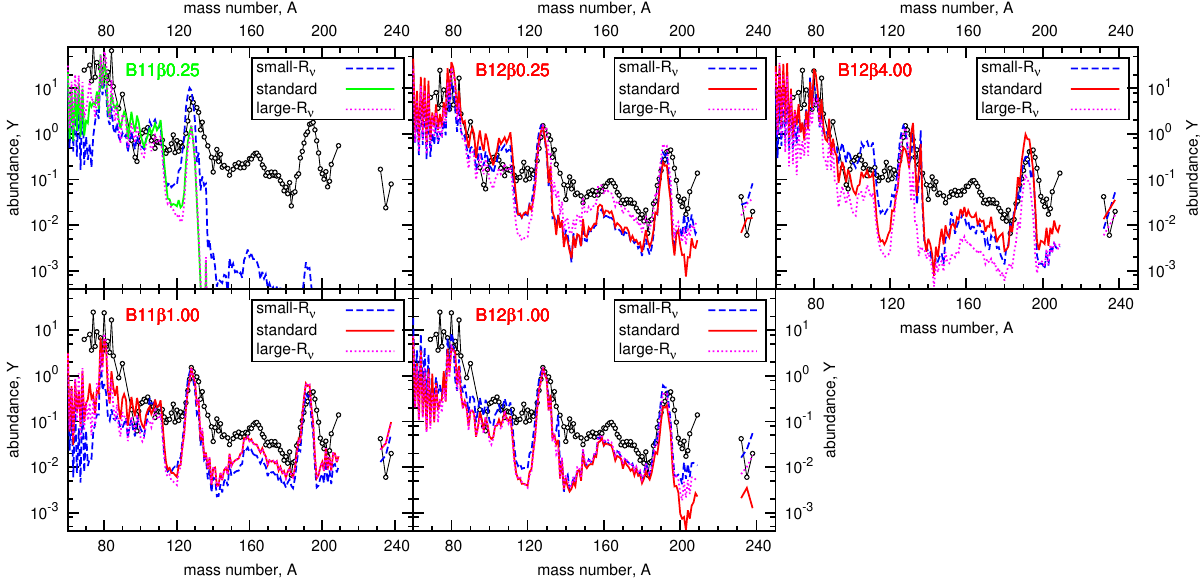}
		\caption{The final abundances ($Y$) as a function of mass number ($A$),
                based on the different set of ${Y_{\rm e}}$,
		with the solar $r$-process pattern (open dots).
		Each explosion model has three different lines,
                following distribution of ${Y_{\rm e}}$ shown in Figure~\ref{fig-ye-uncertainty},
		which are the results of standard (green for the delayed model or red for the prompt model),
                small-$R_{\nu}$ (blue), and large-$R_{\nu}$ (magenta), respectively.}
	\label{fig-final-abund-uncertainty}
	\end{center}
\end{figure*}

\subsubsection{Physical uncertainties}
\label{sec-uncertainty}

The calculated $r$-process yields have uncertainty due to the nuclear physics input for the reaction network.
As described in the previous section, the prompt-jet reproduces solar $r$-process abundances well,
while there there exists deviations around the second and the third peak
and an underproduction of the rare-earth peak region (Figure~\ref{fig-final-sol}).
It has been pointed out that the deficiency is mostly caused by nuclear physics uncertainty
\citep[compare with, e.g.,][utilizing the same FRDM (1995) mass model]{2012ApJ...750L..22W}.
Improvement of this mass model,
which is a new FRDM \citep[see,][]{2012PhRvL.108e2501M, 2014ApJ...792....6K},
or different mass models \citep[e.g., HFB;][]{2009PhRvL.102o2503G, 2010PhRvC..82c5804G}
are expected to solve this problem.
Resolving these nuclear physics uncertainties is important,
however, as long as we focus on global features of abundance patterns
(e.g., the relative amounts of the second and third peaks and the heaviest nuclei produced $r$-processes),
the difference due to explosion dynamics is much larger.
Therefore, the current discussion about the $r$-process in the MR-SN models
(i.e., the impact of explosion dynamics on nucleosynthesis),
is expected to be applicable to the results based on different reaction networks,
as demonstrated in previous studies \citep[see, e.g.,][]{2006ApJ...642..410N, 2008ApJ...680.1350F}.

Aside from the nuclear physics inputs of the nuclear reaction network,
the physics of PNSs have additional uncertainties.
The evolution of neutron-richness (or $Y_{\rm e}$) during the core-collapse, the core-bounce,
and an early phase of ejection
is influenced by weak interactions.
As shown in Figure~\ref{fig-ye-uncertainty},
the $Y_{\rm e}$ of ejecta has a significant change
within a reasonable range of variation for the radius of a PNS,
adopting $\pm 20\%$ of the standard value.
Following this evaluation, we recalculate $r$-process nucleosynthesis,
of which the results are shown in Figure~\ref{fig-final-abund-uncertainty}.
Each explosion model has a final abundance of the ``standard $Y_{\rm e}$''
(the case of high resolution in Figure~\ref{fig-final-abund}),
with different sets of $Y_{\rm e}$'s that correspond to
the small-$R_\nu$ and large-$R_\nu$ (described in Section~\ref{sec-ye}), respectively.

For prompt-jet models
({\tt B11$\beta$1.00}, {\tt B12$\beta$0.25}, {\tt B12$\beta$1.00}, and {\tt B12$\beta$4.00}),
the global feature of the abundance patterns remains,
although the effects of the $Y_{\rm e}$ variation on the final abundances appear
in the amount of the rare-Earth peak, actinides, and digs around the second and third peaks.
We should recall that these isotopes are also sensitive
to nuclear physics inputs for the reaction network calculations.
Therefore, we can argue that the prompt-jet of MR-SNe produces and ejects heavy $r$-process nuclei,
reproducing the solar-abundance curve within the uncertainty of $Y_{\rm e}$.
However, in order to determine the more precise value of $Y_{\rm e}$ and the final abundances,
we need further studies, based on more sophisticated treatment of the neutrino transport.

On the other hand, the delayed-jet {\tt B11$\beta$0.25}
has larger effects on the uncertainty of the PNS on the final products.
This is because of a larger variation of $Y_{\rm e}$ compared with prompt models
due to neutrino absorption.
This is consistent with our discussion on the progress of the $r$-process in Section~\ref{sec-ye},
only focusing on the neutron-richness of tracer particles.
The abundance pattern of the delayed model reaches up to $A \sim 130$ for the standard case,
so that the qualitative difference in the region of nuclei $A > 130$ appears to be significant.
The case of small-$R_{\nu}$, of which the average value of $Y_{\rm e}$ is low,
slightly produces heavier $r$-process nuclei ($A > 120$),
although this still underproduces compared with the solar abundances.
This difference becomes important when we compare the final abundance
with weak $r$-process patterns in the following section.

\subsubsection{The effect of Jet-deformation}
\label{sec-jet-kink}

%%% Fig. 15
\begin{figure}[htbp]
  \begin{center}
    \includegraphics[width=\hsize]{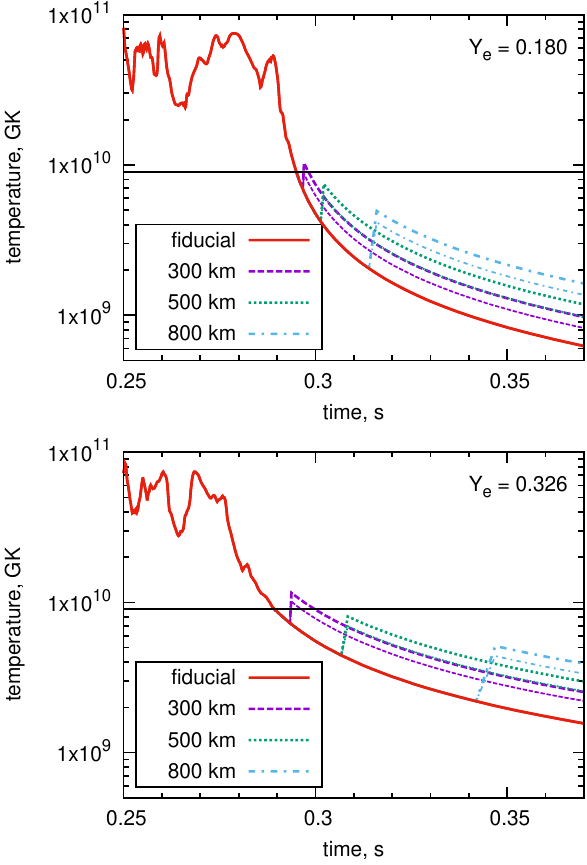}
    \caption{
    The evolution of temperature in GK for selected tracer particles,
    labeled by the value of $Y_{\rm e}$ at the end of NSE.
    Heating due to the magnetic reconnection is taken into account
    at $300$, $500$, and $800$~km from the center, respectively.}
    \label{fig-hydro-kink}
  \end{center}
\end{figure}

MR-SNe models employed in this study are based on 2D axisymmetric simulations,
which ignore several hydrodynamical instabilities that possibly occur in 3D.
In particular, as shown by \cite{2014ApJ...785L..29M},
the launched jet-like can be deformed and destroyed via a hydrodynamical instability
(e.g., the kink instability).\footnote{
This, of course, also depends on the exact rotation rate and magnetic field strength,
see also the 3D simulations of \cite{2012ApJ...750L..22W}.}
The turbulent motion of deformed jets can induce reconnection of the strong magnetic fields
and part of the magnetic energy can be converted to thermal energy.
This heating raises the temperature during the expansion and
can possibly destroy heavy elements that were made in the early jet ejection.
An increase in the initial entropy for the same $Y_{\rm e}$
is not an obstacle for $r$-process production,
in fact, it provides more promising conditions \citep[see,][]{1999ApJ...516..381F}.
But, if such products experience a strong heating after their initial production,
these heavy elements can be destroyed by photo-disintegration.

Here, we examine the effect of jet deformation on the nucleosynthesis,
with a simplified evaluation of the reheating due to reconnection.
We select two different represented trajectories from the prompt-jet model of {\tt B11$\beta$1.00},
whose temperature evolutions are shown in Figure~\ref{fig-hydro-kink}.
The corresponding $Y_{\rm e}$ at the end of the NSE phase are $0.180$ and $0.326$,
leading to the production of heavy $r$-process nuclei,
including the third peak plus lighter elements up to $A \sim 130$, respectively.
For the location of the energy injection by magnetic reconnection,
we chose three different cases at $300$, $500$, and $800$~km
from the center with typical magnetic field strengths
\citep[for details, see][]{2009ApJ...691.1360T}
of $1.0 \times 10^{14}$, $5.0 \times 10^{13}$, and $2.0 \times 10^{13}$~G, respectively.
At the moment of the deformation, we assume that $25$ or $50$ \% of the total magnetic energy
$E_{\rm mag}$ is converted instantaneously to thermal energy, as denoted by $E_{\rm in}$.
Based on the increased total thermal energy, we estimated the increase in temperature
using the Timmes EoS.\footnote{http://cococubed.asu.edu}
For the cases of larger energy deposition (e.g., $E_{\rm in} = 0.5 E_{\rm mag}$ at $800$~km),
the entropies are increased by $4.09$ and $18.5~k_{\rm B}\;{\rm baryon}^{-1}$
for $Y_{\rm e, nse} = 0.180$ and $Y_{\rm e, nse} = 0.326$ trajectories, respectively.

\begin{figure*}[htbp]
  \begin{center}
    \includegraphics[width=\hsize]{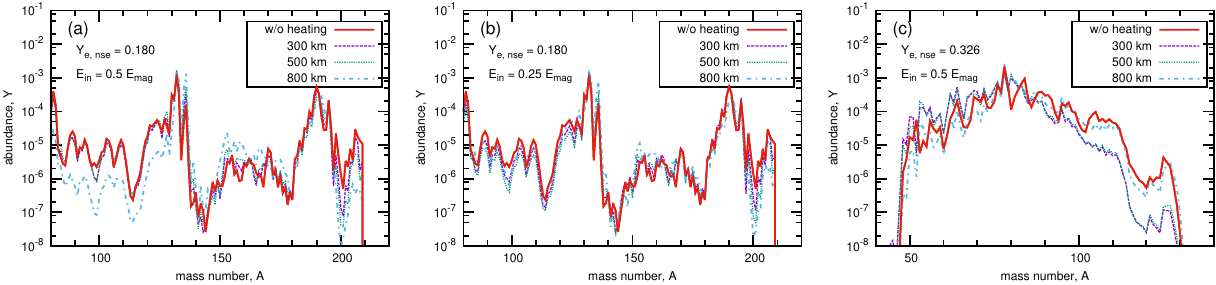}
    \caption{Final abundances of different energy deposition location of the magnetic reconnection
     at $300$, $500$, and $800$~km from the center:
    (a) the trajectory of $Y_{\rm e, nse} = 0.180$ with the energy injection $E_{\rm in} = 0.5 E_{\rm mag}$.
    (b) Same as (a), but the lower injection energy $E_{\rm in} = 0.25 E_{\rm mag}$.
    (c) The trajectory of  $Y_{\rm e, nse} = 0.326$
    with the injection energy of $E_{\rm in} = 0.5 E_{\rm mag}$.}
    \label{fig-fabund-kink}
  \end{center}
\end{figure*}

The results of nucleosynthesis calculations using the modified trajectories
are shown in Figure~\ref{fig-fabund-kink}.
For the cases of $Y_{\rm e} = 0.180$ with larger energy deposition $E_{\rm in} = 0.5 E_{\rm mag}$
(a), the production of heavier nuclei is partially suppressed.
In particular, the model with energy injection at $800$~km shows the largest difference,
changing entire $r$-process nuclei by a factor,
although the main features of the pattern still remain.
For the lower injection energy $E_{\rm in} = 0.25 E_{\rm mag}$ (b), all models show a smaller impact by the heating,
except for the isotopes $A > 200$.
For the $Y_{\rm e} = 0.326$ trajectory with large $E_{\rm in} = 0.5 E_{\rm mag}$,
which produces mostly lighter $r$-process nuclei $A < 130$,
most of the isotopes with $A < 100$ are not changed.
Similar to case (a) above, the heavier nuclei with $A > 100$
are somewhat suppressed by the late temperature increase.

Based on the estimation of the simplified models, we can conclude that
in the case of very strong energy deposition due to magnetic reconnection (e.g., if kink instabilities occur),
the late temperature increase can lead to the destruction of heavy elements after their initial production.
Therefore, the suppression of the heaviest elements can take place (reducing them by a factor),
but these effects do not change the main feature of our current results.

%%% Fig. 16

\subsection{$r$-process Enhanced Metal-poor stars}
\label{sec-weak-r}

\begin{figure}[t]
  \begin{center}
    \includegraphics[width=\hsize]{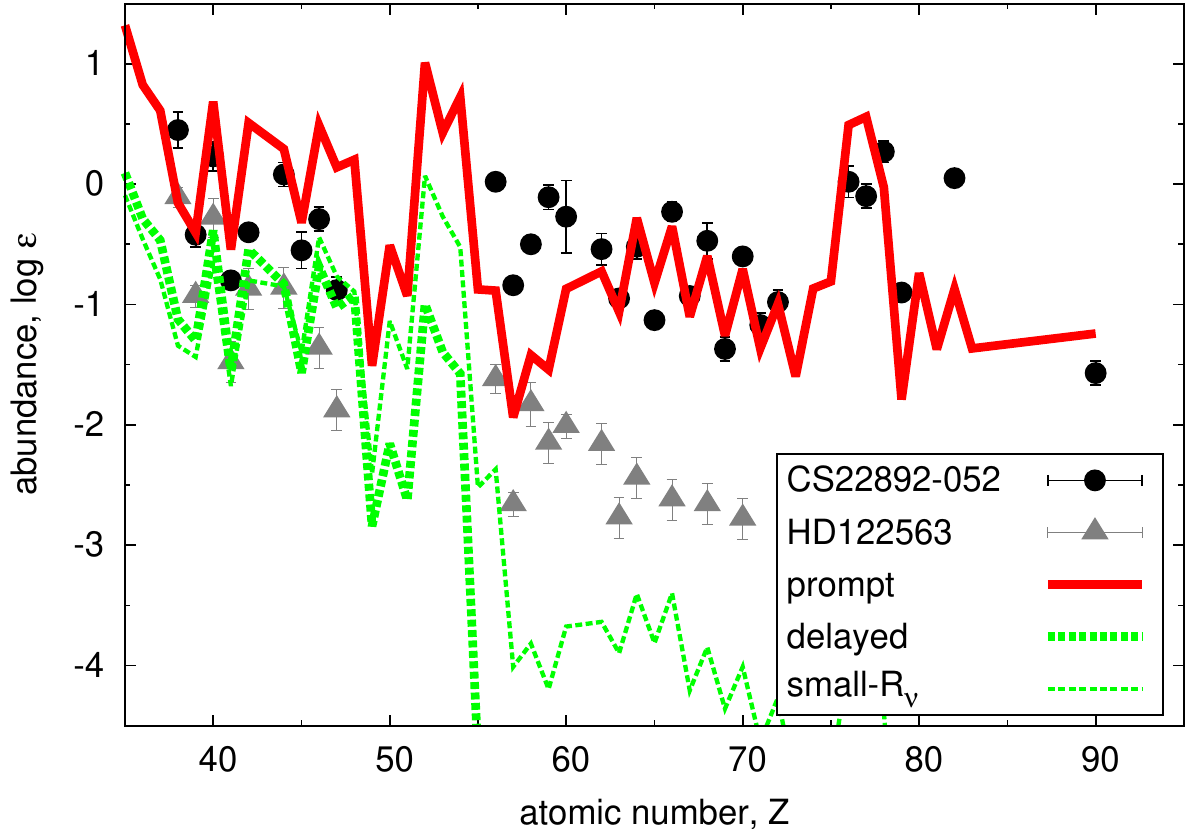}
    \caption{
      Final abundances as a function of atomic number $Z$
      with the abundances of $r$-process-rich metal-poor stars
      of CS22892-052 and HD122563.
      The solid line (red), and the thick dashed line (green), and the thin dashed line (green)
      are a prompt-jet model ({\tt B11$\beta$1.00})
      and a delayed-jet model ({\tt B11$\beta$0.25}) and a delayed model with small-$R_{\nu}$,
      respectively.}
    \label{fig-final-weakr}
  \end{center}
\end{figure}

In addition to the solar abundances,
we compare our results with the observed abundances of metal-poor stars.
In Figure~\ref{fig-final-weakr}, results of the calculations are shown
with the abundance patterns of two $r$-process-rich metal-poor stars.
We choose {\tt B11$\beta$1.00} as a typical prompt-jet model,
while we adopt the standard set of $Y_{\rm e}$ and the case of small-$R_\nu$ for
the delayed-jet model {\tt B11$\beta$0.25}.
These final abundance patterns are based on the same result in Figure~\ref{fig-final-sol},
although we replotted as a function of the atomic number $Z$.
In this figure, these are compared with the abundance patterns of
two different $r$-process-enhanced metal-poor stars,
which we adopt CS22892-052 \citep[e.g.,][]{1996ApJ...467..819S}
and HD122563 \citep{2004ApJS..152..113H, 2004ApJ...607..474H}.
They represent solar-like $r$-process abundances and the ``weak $r$-process'' pattern, respectively.\footnote{
There are still several arguments of whether or not a single event originates this pattern.
In this paper, we mention it as an alternative observational $r$-process signature.}
The solar-like distribution has been observed in several metal-poor stars,
which have a remarkably similar feature, or ``university,''
for heavier $r$-process elements ($A \gtrsim 130$).
So far there have not been as many examples reported as weak $r$-process patterns
\citep[see,][]{2008ARA&A..46..241S}.

The final abundances of prompt-jet models are basically agree
with the observational pattern of CS22892-052,
which is a solar-like $r$-process abundance distribution,
although the underproduction is around $Z=60$.
This is of course same as the case of the comparison
with the solar-abundance pattern, as discussed previously.
In the case of the delayed-jet, on the other hand, the result of standard $Y_{\rm e}$
produces nuclei up to $Z \sim 55$ and has no production for further heavy elements.
As we consider the uncertainty of $Y_{\rm e}$,
this severe underproduction appears to be dissolved,
because the delayed model with smaller $R_{\nu}$ (smaller $Y_{\rm e}$)
produces heavier $r$-process nuclei.
The $r$-process in the low entropy environment is much depend on
the degree of neutron-richness, so that further lower $Y_{\rm e}$ cases,
which are between the prompt and delayed-jet models,
are expected to be reproduce the entire pattern of HD122563.

When we consider MR-SNe as the origin of nucleosynthetic signatures
in $r$-process-rich metal-poor stars,
we can repeat the discussion on the solar abundances.
The ejecta of prompt-jet models can be a source for CS22892-052,
which is a solar-like $r$-process-rich metal-poor star.
Whereas, the weak $r$-process pattern 
seems to indicate an intermediate feature between our prompt and delayed models.
The question then arises whether its features relate to a single/pure process
from one event or whether it is a weighted superposition of different events
\citep[as discussed by][]{2014ApJ...797..123H}.
In future, more extended parameter studies,
we will explore whether the full weak $r$-process pattern can be reproduced in total.
This might offer the chance to explain the scatter and different $r$-process abundance patterns in the early galaxies.

\subsection{Synthesis of Ni isotopes in Jets}
\label{sec-masses}

In addition to $r$-process nuclei, CC-SNe produce a large amount of iron-group isotopes
by explosive nucleosynthesis.
Radioactive nickel isotopes (e.g., $^{56}{\rm Ni}$)
are of particular importance for connecting nucleosynthesis yields of SNe to several optical observations
\cite[see,][and references therein]{2010ApJS..191...66M}.
For instance, emission of $\gamma$-rays from $^{56}{\rm Ni}$
via the chain
$^{56}{\rm Ni} \rightarrow{ }^{56}{\rm Co} \rightarrow { }^{56}{\rm Fe}$
is the dominant energy source of the early phase of SN light curves.
The jet-like explosion of MR-SNe ejects not only very neutron-rich material from the inner region,
but also outer layers of the iron core $Y_{\rm e} > 0.4$ as shown in Figure~\ref{fig-jet-evol}.
Here, we calculated the ejected masses of $^{56}$Ni,  $^{57}$Ni, and $^{58}$Ni
in the primary MHD-driven jet,\footnote{We excluded {\tt B12$\beta$0.25},
which does not yet feature the full explosive nucleosynthesis.
The current explosion models have been geared mainly
to investigate the $r$-process, as determined by the $Y_{\rm e}$ at the early phase of explosions.}
which is summarized in Table~\ref{tab-ni}.
Synthesis of these isotopes takes place in a distant region from the PNS,
where the effect of neutrino absorption is negligible.
In fact, the final abundances of iron-group elements change only a few percent,
based on the uncertainty of $Y_{\rm e}$ discussed in Section~\ref{sec-uncertainty}.

%%%%% Fig. 17

The mass of $^{56}{\rm Ni}$, $M (^{56}{\rm Ni})$, is in the range of
$0.714 \times 10^{-2} M_\odot$ and $4.27 \times 10^{-2} M_\odot$,
which appears to be correlated with the total ejected mass in Table~\ref{tab-rproc}.
Our models have lower values compared with a typical type SN IIp,
SN~1987A, $M(^{56}{\rm Ni}) = 7.1 \times 10^{-2} M_\odot$
and $M(^{57}{\rm Ni}) = 4.1 \times 10^{-3} M_\odot$ \citep[see, e.g.,][]{2014ApJ...792...10S}.
The jet-like explosion of MR-SNe sometimes has been considered to be
the central engine of luminous SNe Ib/c or hypernovae associated with long GRBs.
However, this amount of $^{56}{\rm Ni}$ can hardly explain these luminous events,
which require $M (^{56}{\rm Ni}) > 0.1M_\odot$.
While, $M (^{56}{\rm Ni}) \sim 0.01 M_\odot$ corresponds to
the lower bound of ordinary SNe,
which is in the faint SN branch
\cite[see, Figure~1 of][]{2006NuPhA.777..424N, 2009ARA&A..47...63S, 2009ApJ...692.1131T}.
This indicates that the delayed-jet model,
of which $^{56}{\rm Ni}$ mass is $\sim 0.01 M_\odot$ with a less energetic explosion,
may be classified in the faint SN.

%%% table 6
\begin{table}[tbp]
\centering
\caption{Masses of nickel isotopes}
\begin{center}
\begin{tabular*}{\hsize}{@{\extracolsep{\fill}}lcccccc}
\hline
\hline
Model
& $M(^{56}{\rm Ni})$ & $M(^{57}{\rm Ni})$ & $M(^{58}{\rm Ni})$ \\
& ($10^{-2} M_{\odot}$) & ($10^{-4}M_{\odot}$) & ($10^{-3}$) \\
\hline
    {\tt B11$\beta$0.25} & $1.08$  & $2.51$ & $0.774$ \\
    {\tt B12$\beta$0.25} & $0.714$ & $2.48$ & $3.31$ \\
    {\tt B12$\beta$1.00} & $1.06$  & $2.91$ & $2.57$ \\
    {\tt B12$\beta$4.00} & $4.27$  & $7.81$ & $6.02$ \\
    \hline
\end{tabular*}
\end{center}
\tablecomments
    {The amounts of $^{56}$Ni, $^{57}$Ni, and $^{58}$Ni in the jet-like ejecta,
      as measured in the solar mass ($M_\odot$).
      The model of {\tt B12$\beta$0.25} is excluded
      because the duration of the simulation is insufficient
      for determining these isotopes.}
\label{tab-ni}
\end{table}

Based on the amounts of Ni isotopes, we can quantitatively discuss
the connection between MR-SNe and several high-energy astronomical phenomena.
An X-ray flash event, XRF~060218, that is associated with SN~2006aj (Type Ic)
requires $^{56} {\rm Ni} \sim 0.2 M_\odot $ and $^{58} {\rm Ni} \sim 0.05 M_\odot$
by optical observations
\citep{2006Natur.442.1018M, 2007ApJ...658L...5M, 2007ApJ...661..892M}.
Although our explosion models show lower Ni-isotope ejecta,
they are still suitable for another peculiar type Ib SN, SN~2005bf,
which has intermediate explosion energy between normal type Ib/c SNe and X-ray flashes.
Our results have the amount of Ni isotopes in the range of
$M(^{56}{\rm Ni}) \leq 0.08 M_\odot$ with $\sim 10^{14}$ -- $10^{15}$ G,
which are values for a magnetar formation scenario \citep{2007ApJ...666.1069M}.
Additionally, the lower ejected mass of $M(^{56}{\rm Ni}) = 0.003 M_\odot$
is suggested for jet-like explosion of SN~2010jp \citep{2012MNRAS.420.1135S},
which may correspond to a less energetic MR-SN models.\footnote
{The collapse model also may be suitable for such low $^{56}$Ni ejecta
\citep[the lower bound of $^{56}{\rm Ni}$ is $3.7 \times 10^{-4} M_\odot$ by][]{2008ApJ...680.1350F}.
However, it is not clear if BH-driven jets are possible
without pollution of yields by magnetar-driven (MR-SN) jets before BH formation
\citep{2009ApJ...704..354H}.}

However, luminous SNe Ib/c and relevant explosive phenomena including GRBs and XRFs
generally require larger amounts of $^{56}{\rm Ni} > 0.2~M_\odot$
\citep[see, e.g.,][]{2013MNRAS.434.1098C},
which the calculated value of $^{56}{\rm Ni}$ for our explosion models hardly explain.
Here, it should be emphasized that we have only focused on
the production of Ni isotopes in the primary jets.
Thus, we have ignored additional ejecta due to neutrino-heating after jet propagation,
which is discussed by \cite{2014ApJ...784L..10S}.
We can expect significant amounts of Ni and other iron-group isotopes in this neutrino-driven ejecta, 
unless the central PNS immediately collapses to BH.
The current study might give a clue for further investigation using more sophisticated models.

% -----------------------------------------------------------------------------------------------
\section{Summary and conclusions}
\label{sec-summary}

We investigated the $r$-process nucleosynthesis,
focusing on the explosion scenario of MR-SNe.
We found characteristic nucleosynthesis signatures,
which are different from isotopic abundances
in canonical neutrino-heating driven CC-SNe.
The results are summarized as follows:
\begin{enumerate}
\item
  A successful $r$-process, producing the solar $r$-process pattern,
  takes place in the jet-like ejecta of all {\it prompt-magnetic-jet} models,
  which have stronger initial magnetic fields.
  The final abundances are in good agreement with solar $r$-process abundances
  (Figure~\ref{fig-final-sol}) within the remaining physical uncertainties.
  The weaker magnetized explosion,
  categorized as the {\it delayed-magnetic-jet},
  produces matter up to the second peak nuclei ($A \sim 120$).
\item
  The prompt-jet can explain the solar $r$-process abundances and solar-like patterns of metal-poor stars,
  while the delayed-jets possibly contribute to ``weak $r$-process''
  patterns of $r$-process-rich metal-poor stars.
  Additionally, our results predict that intermediate abundance patterns
  between a main (solar-like) $r$-process and a weak $r$-process are produced
  by certain conditions of stellar rotation and magnetic fields.
\item
  The total amount of ejecta enriched by $r$-process elements
  is in the range of $\sim 10^{-2} M_{\odot}$ per event.
  This value is about a thousand times larger than
  the typical value of PNS wind ejecta.
  Thus, MR-SNe have a significant contribution
  to the galactic chemical evolution of $r$-process elements, also requiring
  that the frequency of events is relatively low.
  This is consistent with the observational indication of MR-SN events,
  of which frequency is low compared to canonical neutrino-driven SNe.
\item
  The uncertainty of the present $r$-process calculations
  due to the treatment of weak interactions is evaluated,
  focusing on the physics of the PNS.
  Reasonable uncertainties, which allow a $\pm 20\%$ variation
  of the neutrino-sphere radius, modify nucleosynthesis results,
  although the main properties found in this work remain.
  Delayed-jets, which are strongly influenced by neutrino absorption,
  experience larger changes than prompt-jet models.
\item
  We also estimated lower bounds for the amounts of Ni isotopes in jets.
  The amount of $^{56}$Ni ($\sim 0.01 M_\odot$ for our explosion models)
  is significantly lower than the value in the ejecta of canonical CC-SNe ($^{56}{\rm Ni} \sim 0.1 M_\odot$).
  Thus, primary jets of the MR-SN are expected to be found as faint SNe by optical observation.
  However, additional matter can be ejected in a later phase of the explosion,
  which can eject larger amounts of $^{56}$Ni corresponding to luminous hypernovae and/or GRBs.
\end{enumerate}

The prompt-jet models exhibit similar nucleosynthesis features,
in spite of different initial rotation and magnetic fields.
In other words, it is clear that strong polar jet-like explosions, in general,
produce and eject significant amounts of heavy $r$-process material.
This indicates that the production of heavy elements is ``saturated,''
when initial rotation and magnetic fields exceed thresholds.
Although our explosion models are based on the axisymmetric hydrodynamics,
nucleosynthesis properties of the prompt models are expected to be similar
to the result of polar-like jets in full 3D simulations \citep{2012ApJ...750L..22W}.
The impact of the remaining physical uncertainties of the current explosion models
on $Y_{\rm e}$ (and the resulting $r$-process nucleosynthesis) are not negligible,
especially for the delayed-jet explosion, which is more sensitive to neutrino absorption.

The amount of ejected $r$-process-rich matter exceeds $\sim 1 \times 10^{-2} M_\odot$,
which is about a thousand times larger than masses of PNS winds.
Therefore, the upper limit of the frequency of MR-SNe is estimated to be $\sim 0.1 \%$
of all CC-SNe by simple multiplication.
However, we need further observational restrictions to clarify
the role of MR-SN events in the entire galactic chemical evolution.
On the other hand, MR-SNe are more likely to be dominant sites of
$r$-process nucleosynthesis for metal-poor stars
\cite[see, e.g.,][]{2014A&A...565A..51C, 2015MNRAS.452.1970W}.
In the early Galaxy, we expect MR-SNe to play the more significant role in
comparison to other $r$-process events, such as compact binary mergers
(before they set in), and it needs to be investigated whether their frequency is a function of metallicity. 

Additionally, MR-SNe also have an advantage to explain individual observed abundances of metal-poor stars,
like CS22892-052 and HD122563 (discussed in Section~\ref{sec-weak-r}).
Our prompt-jet models reproduce solar-like patterns of metal poor-stars,
while the delayed-jet model contributes a weak $r$-process pattern.
The agreement of the heavy element production for prompt-jets of MR-SNe
corresponds to the observational ``universality'' of heavy $r$-process elements
in metal-poor stars.
On the other hand, the ``diversity'' of $r$-process abundances in metal-poor stars
for lighter $r$-process elements with $A < 120$ appears to have different origins,
and can be interpreted as the outcome of a variety of nucleosynthesis yields in MR-SNe
due to different initial rotation and magnetic fields.
Thus, we expect that MR-SN events contribute to the production of $r$-process nuclei
more actively than other $r$-process sources in the early Galaxy.
Additionally, the result that $r$-process yields of MR-SNe have significant variation
due to the properties of progenitors (i.e., initial rotation and magnetic fields),
is consistent with the need for multiple $r$-process astrophysical sites (different abundance patterns),
which have been suggested by at least two different process
in resent investigations \citep{2014ApJ...797..123H}.

In addition to $r$-process elements, MR-SNe also produce other radioactive nuclei
(e.g., $^{56}$Ni), which are sources of SN light curves.
The amount of $M(^{56}{\rm Ni}) \sim 0.01 M_\odot$ in primary magnetic driven jets
for our explosion models hardly explain the luminosity of SNe Ib/c
or hypernovae associated with long GRBs.
This relatively small amount of $^{56}{\rm Ni}$
corresponds to ``faint SNe'' or could explain magnetar birth events like  SN~2005bf.
Our explosion models eject insufficient amounts of $^{56}{\rm Ni}$ and $^{58}{\rm Ni}$
to explain observations of SN~2006aj associated with XRF~060218.
However, the current explosion models possibly underestimate these amounts,
because we only take into account ejecta in jet-like explosions.
Therefore, further comprehensive studies based on a wider range of parameters
may show MR-SN models that satisfy these values.

The explosion mechanism of MR-SNe has a remaining fundamental problem,
which is the origin of such strong magnetic fields during the core-collapse.
One option is the inheritance from stellar evolution
\citep{2004A&A...422..225M, 2005A&A...440.1041M},
which is a long-standing unsolved problem.
The other option is the magneto-rotational instability (MRI),
which is an enhancement process of magnetic fields in the central core during the collapse and explosion.
The role of MRI during core-collapse has been studied with a focus on several aspects
\citep{2003ApJ...584..954A, 2009A&A...498..241O, 2012ApJ...759..110M, 2014MNRAS.445.3169O, 2015ApJ...798L..22M}.
Recently, global MHD simulations with MRI reported
successful jet-like explosions, starting from moderate magnetic fields of pre-collapse models
\citep{2013ApJ...770L..19S, 2014ApJ...784L..10S}.
Although there are still several problems that should be solved
(i.e., numerical convergence, 3D effects, and initial configurations of magnetic fields),
the effects of the MRI are definitely important for further studies.

A major aspect of the present investigations is the assumption
of initial conditions with fast rotation rates and strong magnetic fields,
which lead to an early and rapid jet formation and explosion.
In reality it might be possible that such strong magnetic fields
do not exist at the onset, and they will only be generated in the PNS
as a result of the MRI or a convective alpha-omega dynamo.
This will happen on a longer timescale, leading initially to conditions like in a neutrino-driven explosion,
which we then expect to be followed by a jet formation due to the amplified magnetic pressure.
In such cases, matter is affected longer by neutrino radiation,
which leads to an increase of $Y_{\rm e}$, and the yields
are expected to be similar to our ``delayed-magnetic-jets,''
unless the launched shock is violently strong enough to push out
very neutron-rich matter from the deeper region of PNS.
The answers to these open questions rely on improved stellar evolution calculations
with clear predictions of rotation and magnetic fields,
as well as long-term MHD simulations in 3D,
following the formation and evolution of the PNS.
Such simulations are also expected to provide us with clues to connect the mechanism of MR-SNe
with the shape and amplitude of the magnetic field in NSs.

The nucleosynthesis properties have been presented in this paper show
that MR-SNe play an important role in galactic chemical evolution
and high-energy astronomical events.
In particular, they may be the dominant source of $r$-process elements in early galaxies.
%The neutron-rich ejecta is expected to be a robust source
%of $r$-process material especially in early galaxies.
%Additionally, MR-SNe are favourable to explain optical observations
%of faint-supernovae and X-ray flashes associate with birth of a magnetar.
The current hydrodynamical simulations, however,
are based on a simplified treatment of neutrino transport.
Further studies with more sophisticated MR-SNe explosion models
will hopefully provide more precise theoretical constraints and and predictions.
Numerical data of trajectories and nucleosynthesis yields used in this study
are available at
\href{http://www.astro.keele.ac.uk/~nobuya/mrsn}{\url{http://www.astro.keele.ac.uk/~nobuya/mrsn}}.

%..............................................................................
\section*{Acknowledgments}

The authors acknowledge the anonymous referee for several useful comments.
They also thank A. Cristini for proofreading the manuscript.
N.N. would like to acknowledge to M. Hashimoto and S. Fujimoto
for continuous support and encouragement.
N.N. also thanks to S.~Wanajo, M.~Ono and H.~Sawai for fruitful discussion.
N.N. was financially supported by the National Astronomical Observatory of Japan (NAOJ)
under the ``FY2012 Visiting Fellowship Program.''
The project was funded by the Swiss National Science Foundation (SNSF)
and the European Research Council under SHYNE (EU-FP7-ERC-2012-St Grant 306901) 
and FISH (ERC-2012-ADG-20120216), and the Grants-in-Aid for the
Scientific Research from the Ministry of Education, Science and Culture of Japan (No. 26870823).
Numerical computations performed in the present study
were in part carried out on computer facilities at CfCA in NAOJ, YITP in Kyoto University,
and the COSMOS at DAMTP, University of Cambridge, under the STFC DiRAC HPC Facility.

\bibliography{ref.bib}

\end{document}